\documentclass[fleqn,usenatbib]{mnras}

\usepackage{newtxtext,newtxmath}

\usepackage[T1]{fontenc}

\DeclareRobustCommand{\VAN}[3]{#2}
\let\VANthebibliography\thebibliography
\def\thebibliography{\DeclareRobustCommand{\VAN}[3]{##3}\VANthebibliography}

\usepackage{graphicx}	
\usepackage{amsmath}	
\usepackage{upgreek}
\usepackage{bm}
\usepackage{cleveref}

\newcommand{\darkemu}{\texttt{darkemu}}
\newcommand{\xihm}{\xi_\mathrm{hm}}
\newcommand{\xihh}{\xi_\mathrm{hh}}
\newcommand{\mr}[1]{\mathrm{#1}}

\defcitealias{projeff}{SP20}

\title[Cluster cosmology with anisotropic boosts]{Cluster cosmology with anisotropic boosts: Validation of a novel forward modeling analysis and application on SDSS redMaPPer clusters}

\author[Y. Park et al.]{Youngsoo Park,$^{1}$\thanks{E-mail: youngsoo.park@ipmu.jp}
Tomomi Sunayama,$^{2}$
Masahiro Takada,$^{1}$
Yosuke Kobayashi,$^{3,1,7}$
Hironao Miyatake,$^{4,1}$
\newauthor
Surhud More,$^{5,1}$
Takahiro Nishimichi,$^{6,1}$
Sunao Sugiyama$^{1,7}$
\\
$^{1}$Kavli Institute for the Physics and Mathematics of the Universe (WPI), The University of Tokyo Institutes for Advanced Study (UTIAS),\\The University of Tokyo, 5-1-5 Kashiwanoha, Kashiwa-shi, Chiba, 277-8583, Japan\\
$^{2}$Division of Particle and Astrophysical Science, Graduate School of Science, Nagoya University, Nagoya 464-8602, Japan\\
$^{3}$Department of Astronomy/Steward Observatory, University of Arizona, 933 North Cherry Avenue, Tucson, AZ 85721-0065, USA\\
$^{4}$Kobayashi-Maskawa Institute for the Origin of Particles and the Universe (KMI),
Nagoya University, Nagoya, 464-8602, Japan\\
$^{5}$The Inter-University Centre for Astronomy and Astrophysics, Post bag 4, Ganeshkhind, Pune 411007, India\\
$^{6}$Center for Gravitational Physics, Yukawa Institute for Theoretical Physics, Kyoto University, Kyoto 606-8502, Japan\\
$^{7}$Department of Physics, The University of Tokyo, Bunkyo, Tokyo 113-0031, Japan
}

\date{\today}
\pubyear{2021}

\begin{document}
\label{firstpage}
\pagerange{\pageref{firstpage}--\pageref{lastpage}}
\maketitle

\begin{abstract}
We present a novel analysis for cluster cosmology that fully forward models the abundances, weak lensing, and the clustering of galaxy clusters. Our analysis notably includes an empirical model for the anisotropic boosts impacting the lensing and clustering signals of optical clusters. These boosts arise from a preferential selection of clusters surrounded by anisotropic large scale structure, a consequence of the limited discrimination between line-of-sight interlopers and true cluster members offered by photometric surveys.
We validate our analysis via a blind cosmology challenge on mocks, and find that we can obtain tight and unbiased cosmological constraints without informative priors or external calibrations on any of our model parameters. We then apply our analysis on the SDSS redMaPPer clusters, and find results favoring low $\Omega_\mr{m}$ and high $\sigma_8$, combining to yield the lensing strength constraint $S_8 = 0.715_{-0.021}^{+0.024}$. We investigate potential drivers behind these results through a series of post-unblinding tests, noting that our results are consistent with existing cluster cosmology constraints but clearly inconsistent with other CMB/LSS based cosmology results. From these tests, we find hints that a suppression in the cluster lensing signal may be driving our results.
\end{abstract}




\section{Introduction}
\label{sec:intro}

Galaxy clusters have long been promised to become exquisite probes for the cosmic growth of structure and the accelerated expansion of the Universe \citep{2009ApJ...692.1060V,rozo10,allen11,weinberg13,huterersnowmass,dodelsoncv}. As tracers of the highest peaks in the primordial matter density fluctuations, and the most massive halos in the Universe born therefrom \citep{bbks,kravtsovborgani}, clusters probe the high-mass end of the halo mass function, exponentially sensitive to cosmological parameters \citep{2001ApJ...553..545H,limahu}. The caveat underlying this sensitivity, however, has always been the calibration of cluster masses. Mass calibrations connect the theoretical predictions for cluster-sized halos to the catalogs of observed clusters in real data, and the precision and accuracy of these calibrations limit the quality of cosmological constraints achievable via cluster studies. 

This has made optically identified clusters from photometric surveys a sample of particular interest. Photometric surveys allow for uniform and complete observations of clusters \citep{rm1,rm2,rm3,rm4,camira}, and at the same time detect weak lensing signals around clusters by observing the shapes of galaxies residing in the background of clusters. By combining the observed cluster abundances with the halo mass information afforded by the cluster lensing signal \citep{2007arXiv0709.1159J,wtg1,wtg4,simet17,muratasdss,mcclintock18,2019PASJ...71..107M}, a self-contained analysis that constrains cosmology and calibrates cluster masses becomes possible \citep{limahu,takadabridle,rozo10,oguritakada11}. The most recent incarnations of this joint cluster cosmology analysis are \cite{costanzisdss} and \cite{desy1cl}\footnote{\cite{tokrause2} presents another important cosmology result incorporating clusters, but is not directly comparable to this study as it also uses galaxy data \citep[see also][]{salcedo20}.}. These studies calibrated cluster masses using cluster lensing signals, then used the calibrations to simultaneously constrain the cosmology and the mass-observable relation (MOR) using cluster abundances. Interestingly, both results (in particular the latter) found that the resulting cosmological constraints favored lower $\Omega_\mr{m}$ and higher $\sigma_8$ compared against other constraints derived from cosmic microwave background (CMB) or large scale structure (LSS) data. These findings hinted at yet unknown systematic effects, or new physics, surrounding optical clusters.

Simultaneously with these results, we showed in \cite{projeff} (\citetalias{projeff} henceforth) that the interaction between optical cluster finding algorithms and the weaker line-of-sight resolution of photometric surveys resulted in observed cluster samples with a characteristic anisotropy in the LSS around them. Briefly, \citetalias{projeff} populated galaxies into halos from N-body simulations following an HOD prescription, ran a mock cluster finder algorithm on these galaxies to generate catalogs of mock observed clusters, and studied the lensing and clustering signals exhibited by these mock clusters. From these studies, \citetalias{projeff} found that optically identified clusters were likely to be embedded in LSS elongated along the line-of-sight, e.g., filaments aligned with the line-of-sight, and importantly that this anisotropy led to large-scale boosts of the cluster lensing and cluster clustering signals \citep[also see][for a similar discussion]{2018MNRAS.477.2141O}. These \textit{anisotropic boosts}, which have now been observationally confirmed by \cite{tokrause2}, have significant implications for the joint cluster cosmology framework, as they break the isotropic halo model generally assumed in making lensing mass calibrations for clusters. If not properly modeled, these anisotropic boosts inevitably leads to errors in the cluster mass calibrations.

Thus, the natural continuation of these findings from \citetalias{projeff} is to build a novel analysis framework that accounts for the anisotropic boosts in the cluster lensing and cluster clustering signals, and the main goal of this paper is to achieve exactly that. Namely, in this work we build an analysis framework combining cluster abundances, lensing, and clustering, and introduce an explicit model for the anisotropic boosts. We focus on two major research questions:
\begin{enumerate}
    \item Can we obtain unbiased and competitive cosmological constraints from clusters with freely varied systematics models, in particular for the anisotropic boost?
    \item If we can validate this approach, what results do we find when we perform the analysis on real data and how do they compare against existing results?
\end{enumerate}
Let us elaborate further on these two questions. In (i), we are interested in studying the cosmological information content in cluster observables as agnostically as possible, i.e., without informative priors or simulation-based calibrations. We achieve this by using flat uninformative priors for all of our model parameters, making the functional forms of our modeling ingredients the only externally informed elements of our analysis. This approach is complementary to those from \cite{costanzisdss} and \cite{desy1cl}, which involve multiple informative priors and calibrations. As for (ii), we are mainly interested in whether the explicit modeling of anisotropic boosts leads to different cosmological conclusions from existing cluster cosmology results. \cite{desy1cl} argued that their results are most likely impacted by a suppression in the lensing signals of low-richness clusters, which is the opposite of effects that anisotropic boosts would produce. At the same time, it is unclear whether this suppression is limited to the particular choice of data set and modeling, or it globally affects cluster cosmology analyses. We seek to answer this question with a completely independent combination of data and modeling approach.

This paper is structured as follows. In \Cref{sec:mod}, we describe the theoretical model we employ, including emulator-based halo model predictions and models for common cluster systematics as well as for anisotropic boosts. In \Cref{sec:sim}, we discuss tests of this model on unblinded mocks and the subsequent validation of our fiducial analysis via a blind cosmology challenge on an independent set of mocks. We then discuss the blind application of our analysis pipeline on real data in \Cref{sec:sdss}, and present a series of post-unblinding tests in \Cref{sec:post}. We finally summarize our work and discuss its implications in \Cref{sec:conc}.

\section{Modeling}
\label{sec:mod}

\subsection{Emulator predictions}
\label{sec:mod:emu}
Throughout this paper, we use the Dark Emulator \citep[\darkemu ~ henceforth]{darkemu} software to generate base predictions for halo model quantities. Specifically, for a given set of cosmological parameters $\bm\uptheta$ and redshift $z$ there are three fundamental quantities that we obtain from \darkemu:
\begin{itemize}
    \item The halo mass function $\mathrm{d}n/\mathrm{d}\ln M$
    \item The 3D halo-matter correlation function $\xi_\mathrm{hm}(r|M)$
    \item The 3D halo-halo correlation function $\xihh(r|M,M')$
\end{itemize}
with halo masses $M$ and $M'$. Note that we suppress the $\bm\uptheta$ and $z$ dependence of quantities in our notation throughout. The emulator was trained based on the measurement of the above quantities in a suite of 101 numerical simulations that were run to optimally span a 6-parameter space of cosmological models. The accuracy of the emulation was tested with a set of test simulations that also spanned the parameter space and were reserved during the training process. The emulator is accurate to 2\% in the predictions of halo-matter correlation functions relevant for galaxy-galaxy lensing, and to 4\% in the predictions of halo autocorrelation functions relevant for the estimates of clustering, in the range of its validity ($0<z<1.48$ and $M\gtrsim10^{12}h^{-1}M_\odot$). The \texttt{darkemu} code then analytically converts the 3D statistics to the projected 2D statistics, which are the main focus of the current investigation. The emulation accuracy was also tested for the projected lensing and clustering statistics, using mock galaxy catalogs roughly corresponding to the BOSS CMASS galaxy sample \citep{boss}. It was confirmed that the projected statistics from the emulator agreed with measurements from the mock galaxy catalogs at the same level as the original 3D statistics.
In using the emulator to calculate our model predictions, we assume a flat $\Lambda$CDM model of cosmology with the neutrino density fixed to $\Omega_\nu h^2 = 0.00064$, and ignore the redshift evolution of the halo statistics by calculating these quantities at a single representative redshift. We also adopt the ``$200m$'' halo definition throughout, following the \texttt{darkemu} conventions.

We then map these quantities to cluster observables as follows. First, for cluster abundances in a given richness bin, we have
\begin{eqnarray}
    N_{\mathrm{c},i} & = & \Omega \int\! \mathrm{d}z~ \frac{\mathrm{d}^2V}{\mathrm{d}z\mathrm{d}\Omega} \int_{\ln \lambda_\mathrm{i,min}}^{\ln \lambda_\mathrm{i,max}}\! \mathrm{d}\ln \lambda \int\!\mathrm{d} \ln M \frac{\mathrm{d}n}{\mathrm{d}\ln M} P(\ln \lambda|M) \nonumber \\
    & = & \Omega \int\! \mathrm{d}z \frac{\chi^2(z)}{H(z)} \int\! \mathrm{d}\ln M \frac{\mathrm{d}n}{\mathrm{d}\ln M}S_i(M),
\end{eqnarray}
with richness $\lambda$, the effective survey area $\Omega$,
comoving distance $\chi$, and the Hubble constant $H$. 
We follow \cite{muratasdss} to define our mass-richness relation $P(\ln \lambda|M)$ as a log-normal distribution, i.e. $\ln \lambda \sim \mathcal{N}(\mu_\mathrm{M}, \sigma_\mathrm{M}^2)$, with mean $\mu_\mathrm{M}$ and standard deviation $\sigma_\mathrm{M}$ given by
\begin{eqnarray}
    \mu_\mathrm{M} & = & A + B \ln \left( \frac{M}{M_\mathrm{piv}} \right), \\
    \sigma_\mathrm{M} & = & \sigma_0 + q \ln \left( \frac{M}{M_\mathrm{piv}} \right).
\end{eqnarray}
We set our pivot mass $M_\mr{piv} = 3\times10^{14}M_\odot/h$. This then gives the following mass selection function $S_i(M)$,
\begin{eqnarray}
    S_i(M) & = & \int_{\ln \lambda_\mathrm{i,min}}^{\ln \lambda_\mathrm{i,max}} d\ln \lambda ~ P(\ln \lambda|M) \nonumber \\
    & = & \Phi(\ln \lambda_\mathrm{i,max}|\mu_\mathrm{M},\sigma_\mathrm{M}) - \Phi(\ln \lambda_\mathrm{i,min}|\mu_\mathrm{M},\sigma_\mathrm{M}),
\end{eqnarray}
where $\Phi(x|\mu,\sigma)$ is the cumulative distribution function of a Gaussian distribution with mean $\mu$ and standard deviation $\sigma$:
\begin{equation}
    \Phi(x|\mu,\sigma) = \frac{1}{2}\left[1+\mathrm{erf}\left(\frac{x-\mu}{\sqrt{2}\sigma}\right)\right],
\end{equation}
with $\mathrm{erf}(x)$ being the error function.

Next, we model the cluster lensing signal as follows. We first obtain the cluster-matter correlation function for the $i$-th bin as
\begin{equation}
    \xi_{\mathrm{cm}, i}(r) = \frac{1}{n_{\mathrm{c},i}}
    \int\! \mathrm{d}\ln M~ \frac{\mathrm{d}n}{\mathrm{d}\ln M} S_i(M) \xihm(r|M),
\end{equation}
with the cluster number density for the $i$-th bin $n_{\mr{c},i}$ given by
\begin{equation}
    n_{\mr{c},i} =  \int\! \mathrm{d}\ln M~ \frac{\mathrm{d}n}{\mathrm{d}\ln M} S_i(M).
\end{equation}
We then obtain the surface density of matter around clusters $\Sigma(R)$  as
\begin{equation}
    \Sigma_i(R) = \bar{\rho}_{\mathrm{m},0} \int_{-\infty}^{\infty}\!\mathrm{d}\pi ~ \xi_{\mathrm{cm}, i}\left(\sqrt{R^2+\pi^2}\right),
\end{equation}
where we only consider the matter density in the Universe that contributes to lensing. Here, $R$ is the separation perpendicular to the line-of-sight, and $\pi$ is the separation parallel. We finally obtain the cluster lensing observable, i.e., the excess surface density $\Delta\Sigma(R)$, with the usual formalism
\begin{eqnarray}
    \Delta\Sigma_i(R) & \equiv & \left<\Sigma_i\right>(<R)-\Sigma_{i}(R) \nonumber\\
    & = &  \frac{2}{R^2} \int_0^R \mathrm{d}R' R' \Sigma_i(R') - \Sigma_i(R).   
\end{eqnarray}
Equivalently, $\Delta\Sigma(R)$ is given by
\begin{eqnarray}
\Delta\Sigma(R) & = & 
\bar{\rho}_{\mathrm{m},0} \int\! \frac{KdK}{2\pi} J_2(KR)C_\mr{cm}(K) \\
& \simeq & \bar{\rho}_{\mathrm{m},0} \int\! \frac{kdk}{2\pi} J_2(kR) P_\mr{cm}(k),
\end{eqnarray}
where $C_\mr{cm}(K)$ and $P_\mr{cm}(k)$ are the 2D and 3D cluster-matter power spectra with the 2D wavenumber $K$ and the 3D wavenumber $k$, and $J_2$ is the second-order Bessel function. The last line uses the the Limber approximation \citep{limber,limberkaiser}, which becomes exact under our assumption of a single redshift value
\citep[also see][]{2013MNRAS.435.2345H,2021arXiv211102419M}.

Lastly, we model the cluster clustering signal. Similar to the lensing case, we first derive the cluster autocorrelation function for the $i$-th bin as
\begin{equation}
\begin{split}
    \xi_{\mathrm{cc},i}(r) =   \frac{1}{(n_{\mathrm{c},i})^2}
    \int\!\mathrm{d}M \int\!\mathrm{d}M'~   & \frac{\mathrm{d}n}{\mathrm{d}\ln M} 
    \frac{\mathrm{d}n}{\mathrm{d}\ln M' }  ~ \times \\
    & S_i(M) S_i(M') \xihh(r|M,M').
\end{split}
\end{equation}
We then model the cluster clustering observable, i.e., the projected cluster autocorrelation function $w_\mr{p}(R)$, as
\begin{equation}
    w_\mr{p}(R) = \int_{-\pi_\mathrm{max}}^{\pi_\mathrm{max}}\!\mathrm{d}\pi ~  \xi_{\mathrm{cc},i}\left(\sqrt{R^2+\pi^2}\right),
\label{eq:xicctowp}
\end{equation}
where $\pi_\mr{max}$ is the maximum line-of-sight projection width. We decide to use $\pi_{\rm max}=100\,h^{-1}{\rm Mpc}$ as our fiducial choice after tests in \Cref{sec:sim:test}.

\subsection{Anisotropic boosts}
\label{sec:mod:boost}
\begin{figure}
	\includegraphics[width=\columnwidth]{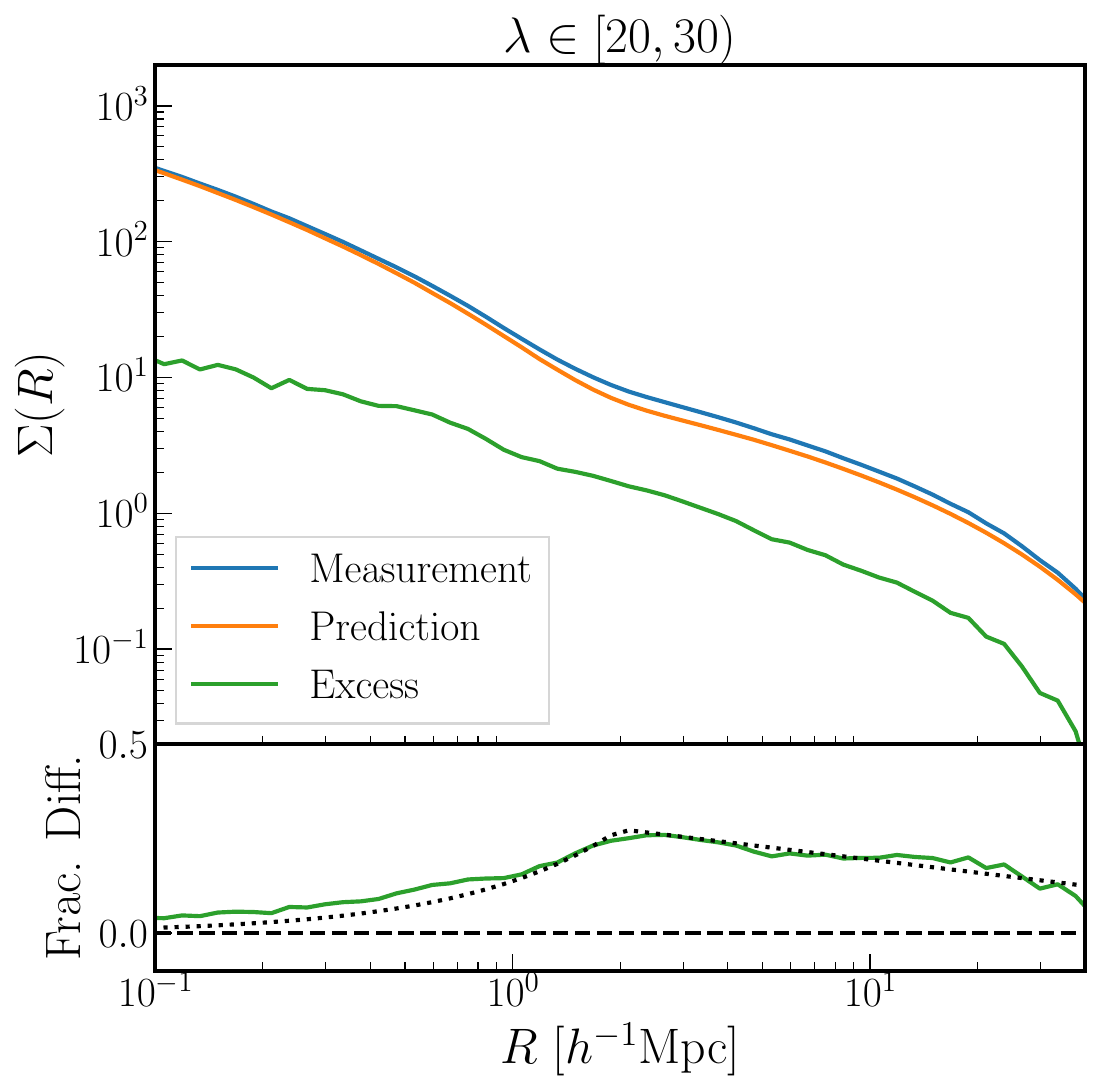}
    \caption{The surface density profiles related to the mock observed clusters  from \citetalias{projeff} simulations. \textit{Top:} the measured surface density (blue) shows an excess (green) when compared against emulator predictions corresponding to the underlying masses (orange). \textit{Bottom:} the excess is shown as a fraction of the emulator prediction. A rough fit using the model in \Cref{eq:anisoboost} (dotted line) is also shown to guide the eye.}
    \label{fig:sigma_proj}
\end{figure}
Our analysis explicitly accounts for the anisotropic boosts on the cluster lensing and cluster clustering signals, reflecting the following insights we first discussed in \citetalias{projeff}:
\begin{itemize}
    \item The LSS around optically identified clusters is not isotropic. Rather, some optical clusters are likely to be embedded within anisotropic LSS, e.g., filaments that are aligned with the line-of-sight.
    \item The most salient impact of the anisotropic LSS is a boost on the large-scale regimes of the cluster lensing and cluster clustering signals.
    \item The respective multiplicative boosts for the lensing and clustering signals show a consistent factor-of-two relation, with the latter being twice as big as the former. This suggests that modifications to the cluster bias may consistently explain both boosts.
\end{itemize}
Drawing on these insights, we model the boost induced by the anisotropic LSS as a scale-dependent multiplicative factor $\Pi(R)$ applied on the cluster bias. This then transforms the isotropic predictions from \Cref{sec:mod:emu} as
\begin{eqnarray}
    \Sigma(R) & = & \Pi(R) \Sigma^\mr{iso}(R), \\
    w_\mr{p}(R) & = & \Pi^2(R) w_\mr{p}^\mr{iso}(R).
\end{eqnarray}
Note that while $\Pi(R)$ directly modifies the observable $w_\mr{p}(R)$ on the clustering side, it acts on the underlying physical quantity $\Sigma(R)$ rather than the direct observable $\Delta\Sigma(R)$ on the lensing side. 
This is because $\Sigma(R)$ is the quantity that represents the raw cluster-matter correlation function; by modifying $\Sigma(R)$, we maintain our interpretation of $\Pi(R)$ as an effective modification of the scale-dependent cluster bias.

We then empirically design a functional form for the projection factor $\Pi(R)$. To do so, we go back to the results from \citetalias{projeff} and examine the measured surface density profiles for mock clusters. In \Cref{fig:sigma_proj}, we show an example from those results. From the fractional differences between the isotropic prediction and the mock results in the bottom panel, we see that the anisotropic boost shows two distinct regimes. On small scales, the boost is roughly linear in $R$ (exponential in $\ln R$) up to a turnover separation of a few Mpc$/h$. After the turnover, the boost slowly falls off logarithmically. Thus, we model $\Pi(R)$ as a piecewise function:
\begin{equation}
    \Pi(R) = 
    \begin{cases}
        \Pi_0 (R/R_0) & \text{for } R \leq R_0, \\
        \Pi_0 + c \ln (R/R_0) & \text{for } R > R_0.
    \end{cases}
\label{eq:anisoboost}
\end{equation}
The boost model is given by three model parameters, $\Pi_0, R_0$ and $c$. The parameters $\Pi_0$ and $R_0$ are defined for each richness bin, while $c$ is shared across all richness bins.
We show a rough fit using this functional form in the bottom panel of \Cref{fig:sigma_proj} to guide the eye.

\subsection{Miscentering}
\label{sec:mod:mis}
Operationally, cluster centers are defined by the location of the brightest cluster galaxy (BCG) within each cluster. This, however, does not always match with the true center of the dark matter halo hosting the cluster \citep{oguritakada11,2013MNRAS.435.2345H,miyatakecl}. When this mismatch happens, the cluster lensing signal measured around the BCG will receive a characteristic dilution on small scales. To model this dilution, we assume that a fraction ($f_\mr{mis}$) of clusters are miscentered by a random variable $R'$ that follows the Rayleigh distribution
\begin{equation}
    P_\mr{mis}(R'|R_\mr{mis}) = \frac{R'}{R_\mr{mis}^2} \exp \left( - \frac{R'^2}{2R_\mr{mis}^2}\right).
\label{eq:miscen}
\end{equation}
As shown by \cite{oguritakada11}, this assumption nicely simplifies the calculation of the averaged lensing profile for the miscentered population in Fourier space:
\begin{eqnarray}
    C_\mr{cm}^\mr{mis} (K) & = & C_\mr{cm}(K) \int\!\mathrm{d}R' J_0 (KR') P_\mr{mis}(R'|R_\mr{mis}) \nonumber \\
    & = & C_\mr{cm}(K) \exp \left(-\frac{1}{2}K^2R_\mr{mis}^2\right).
\end{eqnarray}
The global average is then given by
\begin{equation}
    C_\mr{cm}^\mr{tot} (K) = (1-f_\mr{mis}) C_\mr{cm}(K) + f_\mr{mis} C_\mr{cm}^\mr{mis}(K),
\end{equation}
which can be Fourier transformed back to real space to yield the cluster lensing signal with miscentering accounted for.

\subsection{Cluster photo-$z$ scatter}
The photometric redshifts of clusters are generally better characterized than galaxies, but the scatter in the estimated redshifts still produce non-negligible dilutions in cluster clustering measurements. To model this effect, we assume a Gaussian scatter in the inferred line-of-sight positions of clusters, characterized by the standard deviation $\sigma_\mr{ph}$. This scatter is then taken into account in the calculation of the cluster autocorrelation function $w_\mr p(R)$ as a modification to \Cref{eq:xicctowp},
\begin{equation}
    w_\mr{p}(R) = 2\int_0^\infty\! \mathrm{d}\pi ~ S_\mr{ph}(\pi|\pi_\mr{max}, \sigma_\mr{ph}) ~ \xi_{\mathrm{cc},i}\left(\sqrt{R^2+\pi^2}\right),
\end{equation}
where the line-of-sight separation selection function $S_\mr{ph}(\pi, \sigma_\mr{ph})$ is given by 
\begin{equation}
    S_\mr{ph}(\pi | \pi_\mr{max}, \sigma_\mr{ph}) = 
    \Phi(\pi | -\pi_\mr{max}, \sqrt{2}\sigma_\mr{ph}) - 
    \Phi(\pi | \pi_\mr{max}, \sqrt{2}\sigma_\mr{ph})
\end{equation}
with $\Phi(x|\mu,\sigma)$ being the cumulative distribution function for a Gaussian distribution with mean $\mu$ and standard deviation $\sigma$.

Note that we do not consider the impact of photo-$z$ scatter for abundances and lensing. For abundances, as our model does not have a redshift dependent richness-mass relation, the effect of photo-$z$ scatter is essentially a smoothing of the redshift boundaries as we integrate over the entire redshift range. We estimate this effect to be very small -- less than 0.2\% assuming a Planck cosmology and typical cluster photo-$z$ errors. For lensing, photo-$z$ scatter will confound the estimated critical surface density values used to translate tangential shear measurements to $\Delta\Sigma$. We estimate this effect to be less than 0.4\%, also very small.

\subsection{Parameter inference}
Throughout this work, we assume a Gaussian likelihood
\begin{equation}
    \ln P(\mathbf{d}|\bm\uptheta) = 
    -\frac{1}{2} \left[\mathbf{d}-\bm{\upmu}(\bm\uptheta)\right]^\intercal \mathbf{C}^{-1} \left[\mathbf{d}-\bm{\upmu}(\bm\uptheta)\right]
\end{equation}
with measurements $\mathbf{d}$, parameters $\bm\uptheta$, model predictions $\bm\upmu(\bm\uptheta)$, and covariances $\mathbf{C}$. Based on this likelihood, we perform Bayesian parameter inferences for the posterior $P(\bm\uptheta|\mathbf{d})$ as 
\begin{equation}
    P(\bm\uptheta|\mathbf{d}) \propto P(\mathbf{d}|\bm\uptheta) P(\bm\uptheta),
\end{equation}
with our likelihood $P(\mathbf{d}|\bm\uptheta)$ and prior $P(\bm\uptheta)$. We perform our likelihood analysis with the \texttt{cosmosis} \citep{cosmosis} software\footnote{\url{https://github.com/joezuntz/cosmosis}}, using the \texttt{polychord} \citep{polychord1, polychord2} and the \texttt{multinest} \citep{multinest1,multinest2,multinest3} samplers integrated therein to sample our posteriors.  We also use the \texttt{chainconsumer} \citep{chainconsumer} software to plot confidence regions for our posterior samples. Under the fiducial analysis setup we describe further in \Cref{sec:sim:test}, we end up with a total of 27 independently varied model parameters: 5 for cosmology, 4 for MOR, 9 for anisotropic boosts, 8 for miscentering, and 1 for the cluster photo-$z$ scatter.

\section{Tests and Validation}
\label{sec:sim}
\subsection{Model tests on \citetalias{projeff} mocks}
\label{sec:sim:test}

We first test our model in \Cref{sec:mod} using the unblinded mock cluster catalogs and measurements from \citetalias{projeff}. Briefly, these mock catalogs start with N-body simulations and resulting halo catalogs from \cite{darkemu}. Identified halos with masses greater than $10^{12}M_\odot/h$ are then populated with galaxies, following an HOD prescription for the number of galaxies populated and a Navarro-Frenk-White \citep[NFW;][]{nfw} profile for the spatial distribution of populated galaxies. Finally, a cluster finder algorithm based on \cite{sunayamamore} is run on the galaxy catalog, yielding the catalog of mock-observed clusters. We use a total of 19 realizations of this mock catalog, each with $1~ (\mathrm{Gpc}/h)^3$ volume and all at redshift $z=0.251$ to emulate the SDSS redMaPPer catalog, and additionally 14 realizations of $2 ~(\mathrm{Gpc}/h)^3$ volume snapshots for better statistics on the cluster clustering measurements. For more details on the test simulations, as well as the measurement procedures we use on the mock catalogs, we refer the reader to Sections~2.2 -- 2.5 of \citetalias{projeff}.

\begin{table}
	\centering
	\caption{The model parameters varied in our fiducial analysis and the respective priors assumed.}
	\label{tab:priors}
	\begin{tabular}{cc} 
		\hline\hline
		Parameter & Prior \\
		\hline
		\multicolumn{2}{c}{\bf Cosmological parameters}\\
		$\Omega_\mr{b} h^2$ & $\mathcal{U}(0.0211375, 0.0233625)$\\
		$\Omega_\mr{c} h^2$ & $\mathcal{U}(0.10782, 0.13178)$\\
		$\ln 10^{10} A_\mr{s}$ & $\mathcal{U}(2.5,3.7)$\\
		$n_\mr{s}$ & $\mathcal{U}(0.92, 1.01)$\\
		$\Omega_\Lambda$ & $\mathcal{U}(0.54752, 0.82128)$\\
		\hline
		\multicolumn{2}{c}{\bf Mass-richness relation}\\
        $A$ & $\mathcal{U}(0.5, 5.0)$\\
        $B$ & $\mathcal{U}(0.0, 2.0)$\\
        $\sigma_0$ & $\mathcal{U}(0.0, 1.5)$\\
        $q$ & $\mathcal{U}(-1.5, 1.5)$\\
        \hline
        \multicolumn{2}{c}{\bf Anisotropic boost parameters}\\
        $\Pi_\mathrm{0,1-4}$ & $\mathcal{U}(0.0, 1.0)$ \\
        $R_\mathrm{0, 1-4}$ & $\mathcal{U}(0.1, 5.0)$ \\
        $c$ & $\mathcal{U}(-2.0, 2.0)$\\
        \hline
		\multicolumn{2}{c}{\bf Miscentering}\\
		$f_\mr{mis,1-4}$ & $\mathcal{U}(0.0, 1.0)$\\
		$R_\mr{mis,1-4}$ & $\mathcal{U}(0.0, 1.0)$\\
        \hline
		\multicolumn{2}{c}{\bf Cluster photo-$z$ scatter}\\
		$\sigma_\mr{ph}$ & $\mathcal{U}(1.0, 100.0)$\\
		\hline \hline
	\end{tabular}
\end{table}

Our goal in this step is to use these unblinded mock catalogs to identify the optimal form of the analysis. We begin by fixing the basic format of our data vector. The mock catalog is divided into four richness bins, respectively defined as $\lambda \in [20,30), [30,40), [40,55),$ and $[55,200)$. Our data vector is then given by
\begin{equation}
    \mathbf{d} = \left\{ \mathbf{N}_\mathrm{c}, \bm{\upDelta\upSigma}_1, \bm{\upDelta\upSigma}_2, \bm{\upDelta\upSigma}_3, \bm{\upDelta\upSigma}_4, \mathbf{w}_\mr{p,1}, \mathbf{w}_\mr{p,2}, \mathbf{w}_\mr{p,3}, \mathbf{w}_\mr{p,4} \right\},
\end{equation}
where $\bm{\upDelta\upSigma}_i$ and $\mathbf{w}_\mr{p,i}$ respectively represent the cluster lensing and cluster clustering signals for the $i$-th richness bin, and $\mathbf{N}_\mathrm{c}=\{N_\mathrm{c,i}\}$ is the vector of cluster counts for the four richness bins. Since the original measurements in \citetalias{projeff} assume perfect knowledge of cluster positions, in this work we add more reality to these measurements by introducing miscentering and photo-$z$ errors. For miscentering, we randomly displace the cluster centers in the 2D plane perpendicular to the line-of-sight, following the same model in \Cref{sec:mod:mis}. For photo-$z$ errors, we again randomly displace the cluster centers but now in the line-of-sight direction, following a Gaussian distribution.

Starting from this setup, we test a number of analysis choices including:
\begin{itemize}
    \item Range and binning of radial scales to be used for analysis,
    \item The line-of-sight projection width ($\pi_\mr{max}$) for $w_\mr p$,
    \item Functional form of the effective bias $\Pi(R)$ modeling the anisotropic boosts on lensing and clustering observables,
    \item Dependence on priors and parameter space volumes.
\end{itemize}
Each test takes the form of a likelihood analysis, and our key concern is the robust recovery of the cosmological parameters $\Omega_\mr{m}$ and $\sigma_8$, and their combination $S_8 = \sigma_8 (\Omega_\mr{m}/0.3)^{0.5}$. These parameters are derived from the five input cosmological parameters for \texttt{darkemu} ($\Omega_\mr{b}h^2$, $\Omega_\mr{c}h^2$, $\ln 10^{10} A_\mr{s}$, $n_\mr{s}$, and $\Omega_\Lambda$) that we directly vary.
As mentioned above, there are three versions of measurements that we consider: \texttt{nomis-test}, the original \citetalias{projeff} measurements without miscentering or photo-$z$ scatter; \texttt{mis-test}, the version with miscentering included; and \texttt{photo-test}, the version with miscentering and photo-$z$ scatter included. Each version of measurement is then analyzed with a corresponding model, i.e., a model that takes into account the set of systematic effects included. 
We use covariances estimated directly from the SDSS redMaPPer catalog, version 6.3, via jackknife resampling.
The resampling uses 83 independent jackknife regions, and we assume a block-diagonal covariance matrix where each of the 9 components of our data vector are considered uncorrelated. This approximation was shown to be reasonable in previous studies \citep[e.g.,][]{simet17}, and we also find that the off-block-diagonal components of the full covariance matrix are negligible, in part due to the domination of shape noise in the cluster lensing covariances.

\begin{figure}
	\includegraphics[width=\columnwidth]{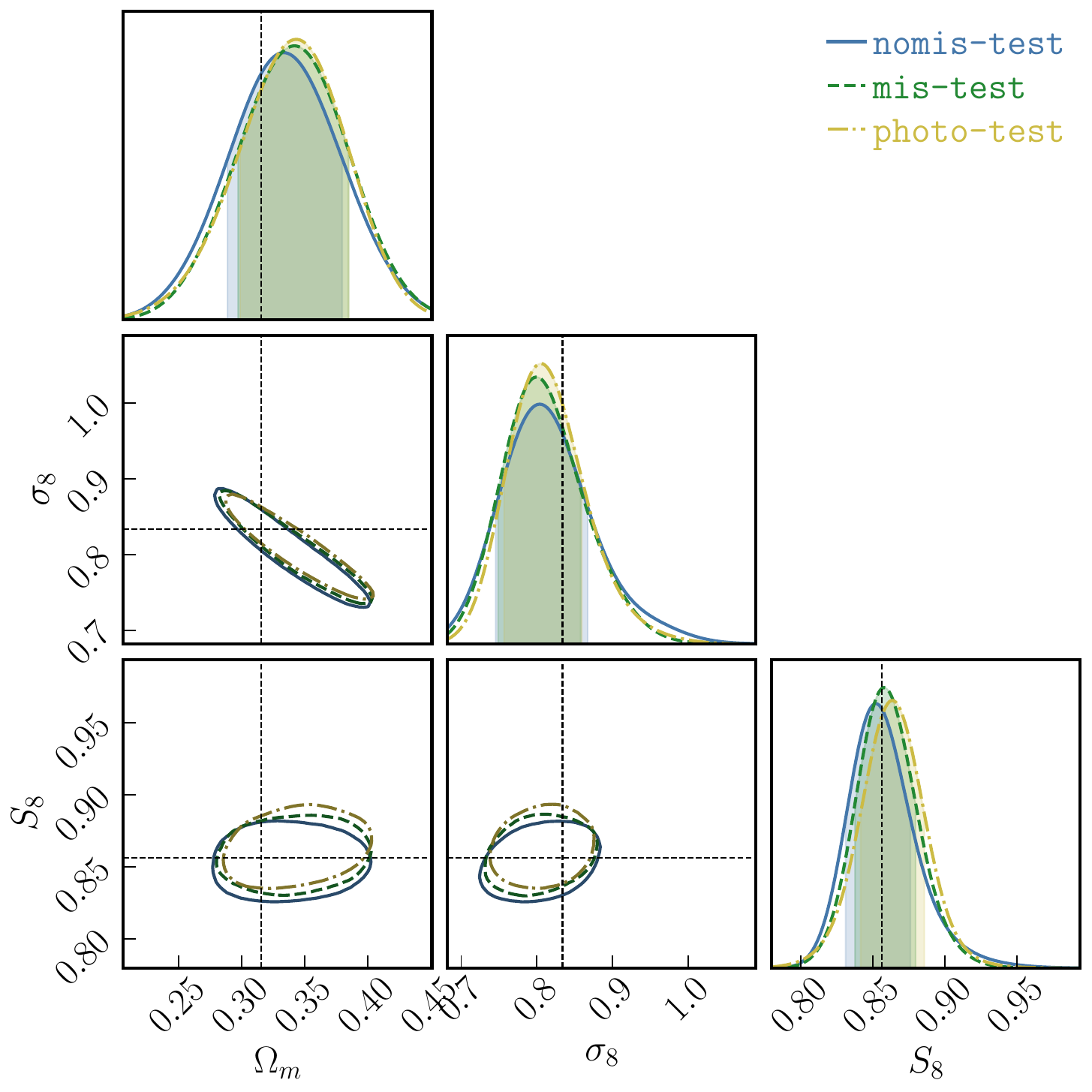}
    \caption{The 68\% confidence regions for cosmological parameters of interest, obtained by running our fiducial pipeline on the test mocks. Blue contours assume no systematics, green assumes the existence of miscentering, and yellow assumes the existence of both miscentering and cluster photo-$z$ scatter. Dashed lines represent the input cosmology for the mocks.}
    \label{fig:test_contours}
\end{figure}

Based on these tests, we define our lensing data vector with 12 logarithmically spaced bins spanning 
$R \in [0.2, 50]$~Mpc/$h$, and our clustering data vector with 6 logarithmically spaced bins spanning $R \in [2,50]$~Mpc/$h$. For the clustering side, we set $\pi_\mr{max}=100$~Mpc/$h$ as our fiducial line-of-sight projection width for $w_\mr{p}$.  Under these analysis choices, we confirm that our anisotropic boost model for $\Pi(R)$ presented in \Cref{sec:mod:boost} yields robust results, with the recovered amplitudes of the anisotropic boosts agreeing well with the values we found in \citetalias{projeff}. We also find that for the \texttt{mis-test} and the \texttt{photo-test} cases our respective constraints on the miscentering and photo-$z$ scatter parameters are in good agreement with the input values. Finally, we find that we recover tight cosmological constraints without significant parameter space volume effects despite the broad, uninformative priors we assume for all model parameters, as detailed in \Cref{tab:priors}.  In \Cref{fig:test_contours}, we show the test results under this fiducial setup. For these results, as well as for the validation results presented in \Cref{sec:sim:val}, we use the \texttt{multinest} sampler. For all three versions of the measurements, we recover the input cosmology well within the 1-$\sigma$ confidence region in the $\Omega_\mr{m}$-$S_8$ plane. Our fiducial model and analysis choices are frozen at this point.
\footnote{After unblinding, we discovered and fixed a minor error relevant for the results in \Cref{sec:sim:test,sec:sim:val}. Fixing the error led to very
small changes in these results, and did not impact our original conclusions.
\Cref{fig:test_contours,fig:challenge_contours} show the results after the fix.}

\subsection{Validation}
\label{sec:sim:val}

With the model and analysis setup successfully developed in \Cref{sec:sim:test}, we move to a blind validation of our model. In this step, an entirely new set of mock cluster catalogs is generated. The new mocks are different from the original \citetalias{projeff} mocks starting from the N-body level; instead of 1 $(\mr{Gpc}/h)^3$ snapshots with $2048^3$ particles, we use 5 independent realizations with 2.5 $(\mr{Gpc}/h)^3$ volumes and $3000^3$ particles, generated with a newly developed TreePM code (T. Nishimichi, S. Tanaka and K. Yoshikawa, in prep). Due to the larger volumes of these new realizations, the mass resolution slightly decreases, which leads to less interloper contamination for detected clusters and thus slightly weaker anisotropic boosts overall.
The underlying cosmological parameters are also meaningfully different from \citetalias{projeff}. In addition, two different miscentering models are used on the mock clusters, one same as in \Cref{sec:mod:mis} and another with a two-population model that assumes a weak but non-negligible degree of miscentering for the previously ``properly centered'' subset of clusters \citep{camira}. We respectively refer to the two resulting versions of measurements as \texttt{mis1-val} and \texttt{mis2-val}. Together with the version without miscentering (\texttt{nomis-val}), this results in three different versions of measurements. All of this information was unknown to the main analyst (YP), except for the changed volumes of the new snapshots necessary for scaling down the cluster counts to 1 $(\mr{Gpc}/h)^3$ equivalent values and the version labels for each measurement.\footnote{We found from \Cref{sec:sim:test} that the photo-$z$ scatter is both well constrained and only weakly impacting for overall constraints, so we choose to focus the validation stress tests on miscentering.}

\begin{figure}
	\includegraphics[width=\columnwidth]{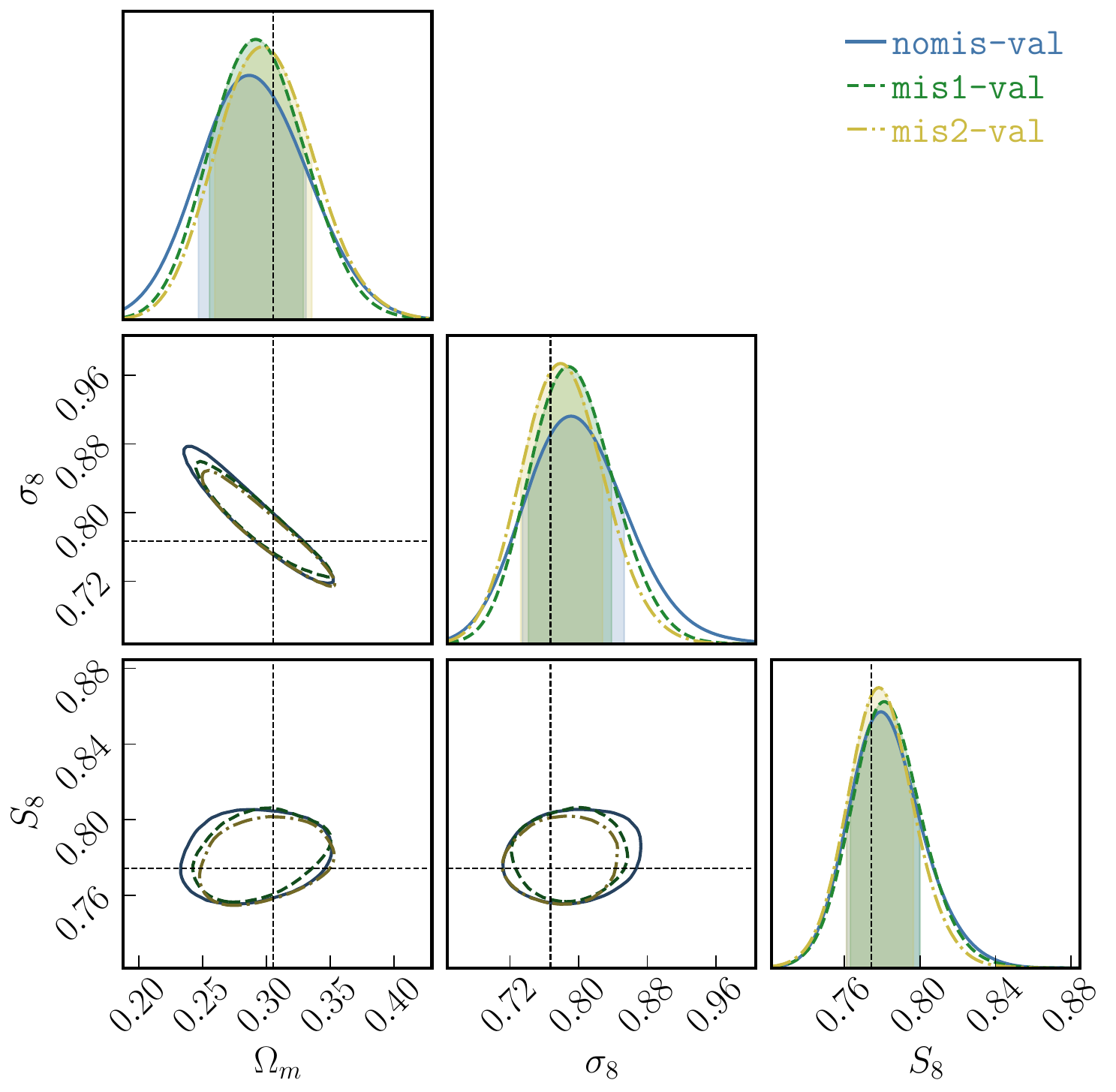}
    \caption{The 68\% confidence regions for cosmological parameters of interest, obtained by running our fiducial pipeline on the validation mocks. The different colored regions represent results from different variations of the validation mock: no miscentering (blue); miscentered according to \Cref{eq:miscen} but with unknown parameters (green); and miscentered with a model different from \Cref{eq:miscen}, again with unknown parameters (yellow). Dashed lines represent the input cosmology for the mocks.}
    \label{fig:challenge_contours}
\end{figure}

We then apply the model and analysis choices from \Cref{sec:sim:test} to analyze the new measurements and test the recovery of input parameters to validate our model. Note that while we use the same $\sim$ 1 $(\mr{Gpc}/h)^3$ equivalent covariances from \Cref{sec:sim:test}, our measurements are averaged over $\sim$ 78 $(\mr{Gpc}/h)^3$. This implies that our validation measurements are nearly noiseless, and consequently that the resulting constraints represent a clean comparison between model-induced biases and the statistical precision we expect from the real data. The validation criterion was determined prior to unblinding as the $1$-$\sigma$ recovery of the input cosmology in the $\Omega_\mr{m}$-$S_8$ plane, i.e., to have the input cosmology within the 68\% confidence contour for the $\Omega_\mr{m}$-$S_8$ plane. In \Cref{fig:challenge_contours}, we show the results from our validation. For all three versions of the validation measurements, our model robustly recovers the input cosmology within the 1-$\sigma$ confidence region in the $\Omega_\mr{m}$-$S_8$ plane. The recovered anisotropic boost parameters were also in agreement with the characteristics of the new validation mocks, showing slightly weaker anisotropic boost amplitudes compared to results from the test mocks. 
Note that we cannot precisely quantify this agreement, as the anisotropic boosts in mocks do not have directly corresponding input parameters. In addition, for the \texttt{mis1-val} case where the input miscentering model and the miscentering model in the analysis match, we again find that our recovered miscentering constraints are in good agreement with the input values. It is also notable that for the \texttt{mis2-val} case, where the input and analysis miscentering models are clearly different, the recovered cosmological constraints are still unbiased and competitive, in fact almost identical to the \texttt{mis1-val} case. This suggests that our assumed miscentering model is flexible enough to accommodate a wider-than-expected range of miscentering in data. The full constraints from our validation analyses are shown in \Cref{app:a}.

The successful validation of our model and analysis choices offers an encouraging answer to the first research question we posed in \Cref{sec:intro}. That is, our analysis enables the combination of cluster abundances, lensing, and clustering to simultaneously achieve tight cosmological constraints and self-calibrations of key systematics without informative priors or calibrations. Compared to previous cluster cosmology studies, our approach adds an additional model dimension of freedom with the anisotropic boosts, but at the same time incorporates a new constraint by including cluster clustering. This could partially explain why our model can successfully obtain tight cosmological constraints despite the added freedom from anisotropic boosts. However, it is still surprising to see that all of our systematics are correctly self-calibrated without priors or external calibrations, in particular for miscentering. Considering that we do not see significant degeneracies between miscentering parameters and cosmology in our posteriors, we postulate that the model behavior induced by miscentering is quite orthogonal to cosmological signatures. This would then imply that even a rough accounting for miscentering, as we do in this work with a simple, freely varied Rayleigh distribution, would successfully mitigate the impact of miscentering on cosmological constraints.

\begin{figure*}
	\includegraphics[width=0.66\columnwidth]{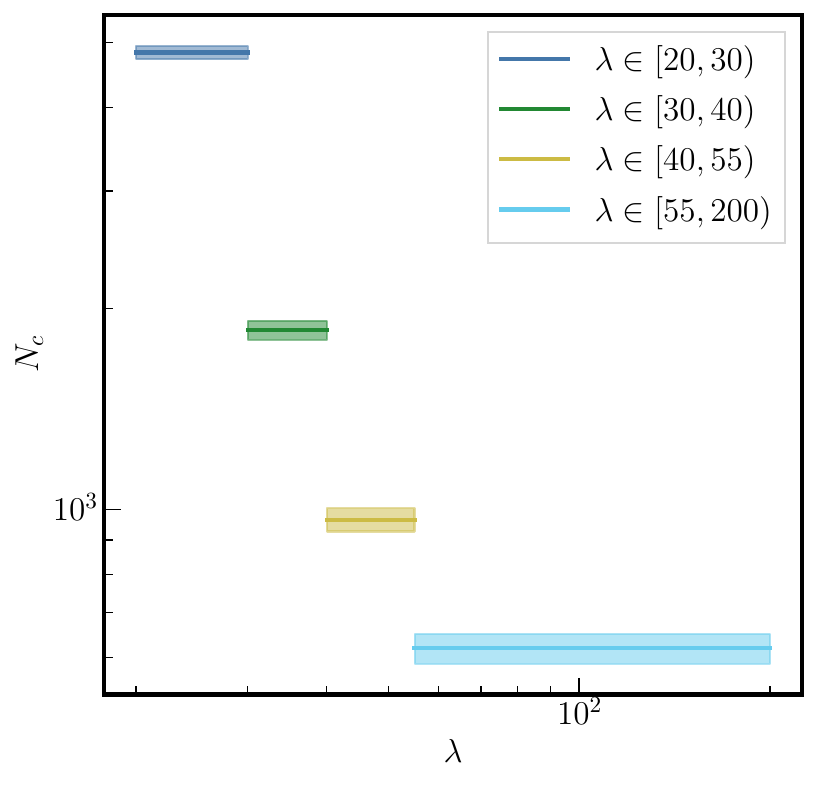}
	\includegraphics[width=0.66\columnwidth]{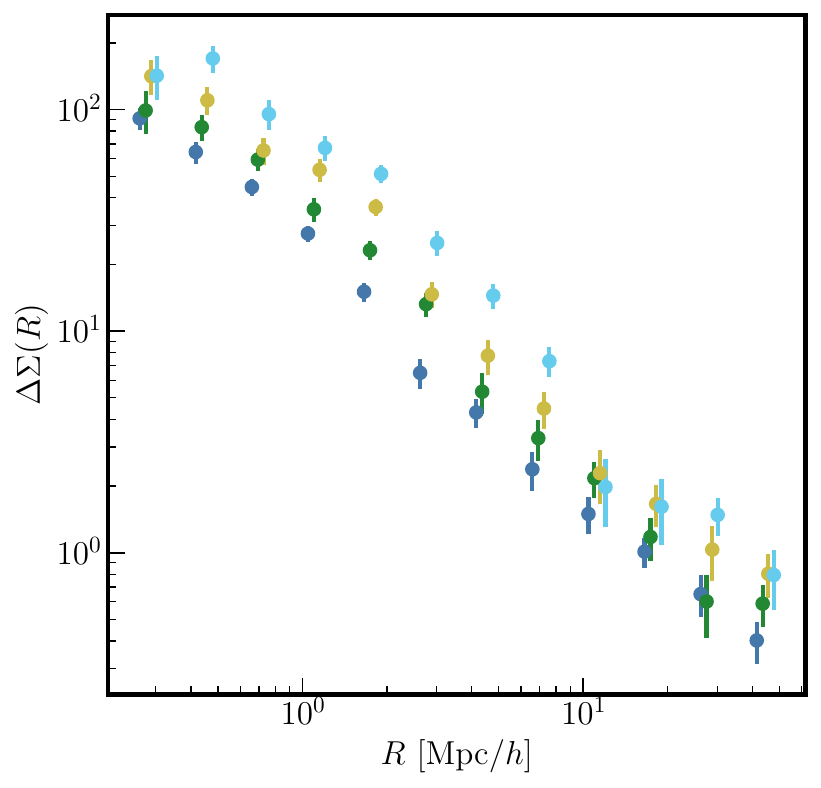}
	\includegraphics[width=0.66\columnwidth]{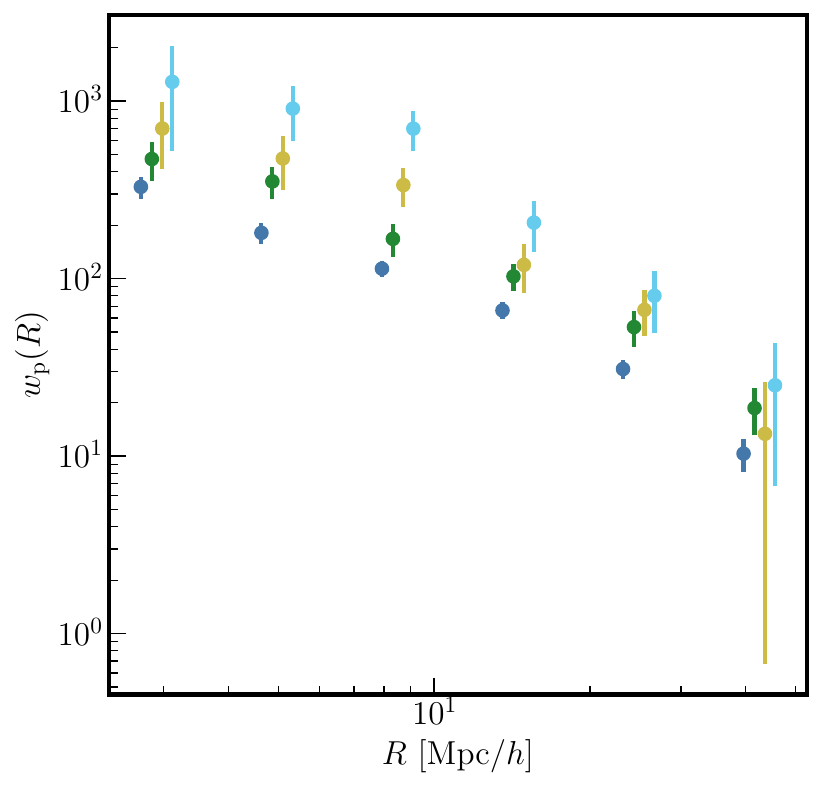}
    \caption{\textit{Left:} Cluster abundance measurements from the SDSS RM catalog. Horizontal bars represent our number counts, and the shaded regions represent the 1-$\sigma$ uncertainties. The colors correspond to our richness bins. \textit{Center:} Cluster lensing measurements and 1-$\sigma$ uncertainties for the four richness bins. Points are offset horizontally for visibility. \textit{Right:} Same as the center panel but for cluster clustering measurements.}
    \label{fig:meas}
\end{figure*}

\begin{figure*}
	\includegraphics[width=2\columnwidth]{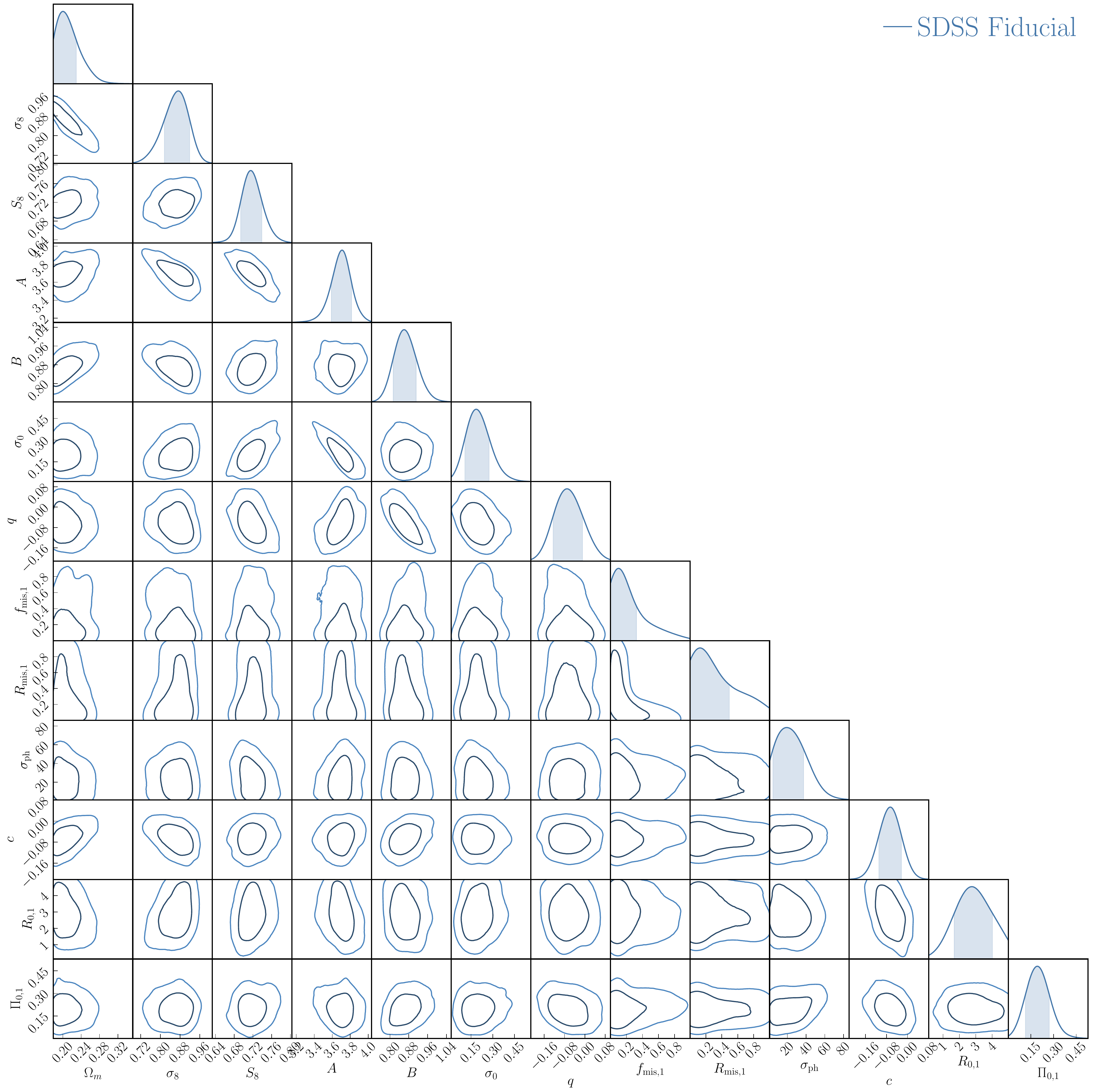}
    \caption{The 68\% and 95\% confidence regions from our fiducial analysis combining SDSS RM cluster abundances, lensing, and clustering. We show here three derived cosmological parameters of interest, four MOR parameters, three systematics parameters for miscentering and cluster photo-$z$ scatter, and three anisotropic boost parameters. For richness bin specific parameters, we only show here results for the first richness bin, and present the rest of the results in \Cref{app:b}.}
    \label{fig:results_fid}
\end{figure*}

\section{Application to SDSS redMaPPer clusters}
\label{sec:sdss}

\subsection{Data}
With our model and analysis choices successfully validated, we now move to apply the framework on real data. We use version 6.3 of the SDSS redMaPPer (RM) cluster catalog \citep{rm1,rm2,rm3,rm4}, based on the SDSS Data Release 8 \citep{sdssdr8} photometry. This catalog covers 10,401 $\mathrm{deg}^2$, and we use a total of 8,379 clusters within the richness range $\lambda \in [20,200)$ and the redshift range $0.1\leq z \leq 0.33$. The redshift range chosen ensures that the RM catalog is very nearly volume-limited. As mentioned in \Cref{sec:mod:emu}, we choose a single representative redshift for this range, $z=0.25$, in calculating model predictions.
For cluster lensing measurements, we follow the methodology of \cite{mandelbaum13} and \cite{miyatakecl}, using the shape catalog from \cite{reyes12}. This shape catalog consists of shapes for 39 million galaxies measured via  the re-Gaussianization technique \citep{hirataseljak}, over a $\sim$9,000 $\mathrm{deg}^2$ overlap with the RM catalog. The shape measurement systematics in this catalog were extensively studied following the procedures defined in \cite{mandelbaum05}, and the image simulations used to calibrate the shape measurements are described in \cite{mandelbaum12}. We use the ZEBRA \citep{zebra, nakajima12} photometric redshifts for these sources.

\subsection{Measurements}
Our analysis consists of three sectors of observables: cluster abundances, cluster lensing, and cluster clustering. In some cases we will respectively refer to these sectors as NC, DS, and WP from here on. To measure cluster abundances, we start with the number counts of clusters in the RM catalog corresponding to each of our richness bins, i.e, for $\lambda \in [20,30), ~[30,40), ~[40,55), ~[55,200)$. The cluster counts, however, are impacted by the detection efficiency of the RM catalog; survey geometries such as masks and boundaries will prevent some clusters from being properly detected. We thus use the richness-dependent detection efficiencies from \cite{muratasdss}, calculated on the exact same catalog as ours, to correct for the missed detections. Also, a background cosmology must be assumed for the following measurement calculations. Under a flat $\Lambda$CDM model the only relevant parameter is $\Omega_\mr{m}$, and we set it to 0.27 in making the measurements. In the subsequent likelihood analyses, we take this assumption into account by introducing a correction to our theoretical predictions, following \cite{more13}.

For lensing, our estimator for $\Delta\Sigma(R)$ is given by
\begin{equation}
\widehat{\Delta \Sigma}(R)=
\frac{\sum_\mr{ls}^{} w_\mr{ls} e_\mr{t,ls}
\Sigma_{\mr{crit,ls}}}
{[1+\hat{m}(R)]\sum_\mr{ls} w_\mr{ls}},
\end{equation}
where the subscripts $\mr l$ and $\mr s$ runs over all lenses and sources considered, $e_\mr{t,ls}$ is the tangential ellipticity for a given lens-source pair,
$\Sigma_{\mr{crit,ls}}$ is the point estimate for the critical surface density corresponding to the lens-source pair given by
\begin{equation}
    \Sigma_{\mr{crit,ls}} = \frac{c^2}{4\pi G (1+z_\mr{l})}
    \frac{\chi_\mr{s}}{\chi_\mr{l} \chi_\mr{ls}},
\end{equation}
with the lens redshift $z_\mr l$, comoving distances to the lens and source $\chi_\mr l$ and $\chi_\mr s$, and the comoving distance between the lens and the source $\chi_\mr{ls}$. The lens-source pair weight $w_\mr{ls}$ is given by
\begin{equation}
w_\mr{ls} = w_\mr{l} w_\mr{s} 
\Sigma_{\mr{crit,ls}}^{-2}
\end{equation}
with the lens and source weights $w_\mr l$ and $w_\mr s$. Note that in our case the lens weights are always unity; the source weights are given by the source catalog. The factor $[1+\hat{m}(R)]^{-1}$ corrects for an overall multiplicative shear bias as a function of $R$ \citep{miller13}. 

We then apply three different corrections to the raw estimated $\Delta\Sigma(R)$. First, we calibrate the measurements for source photo-$z$ biases following Equation 21 of \cite{nakajima12}. This calculation yields a lens redshift dependent bias factor for $\Delta\Sigma$, and we multiply this factor to our measurements. Next, we apply boost factor corrections to account for potential dilutions in the lensing signal from cluster members being mistaken as background sources. The boost factor $\mathcal{B}(R)$ is given by
\begin{equation}
\mathcal{B}(R)=\frac{N_{\mr{r}}}{N_{\mr{l}}}
 \frac{\sum_{\mr{ls}} w_{\mr{l}} w_{\mr{s}}}{\sum_{\mr{rs}} w_{\mr{r}} w_{\mr{s}}},
\end{equation}
where the new subscript $\mr{r}$ refers to the random catalog for RM clusters, with the total number of randoms $N_\mr r$ and weights $w_{\mr{r}}$. The measured signal is then multiplied with $\mathcal{B}(R)$. Finally, we estimate the $\Delta\Sigma(R)$ signal around the randoms and subtract this random signal from our measurements. The random signal detects coherent tangential shear around random points, which should vanish under perfect conditions, and removing it helps account for residual errors in the shape measurements and calibrations.

For the final sector of cluster clustering, we use the Landy-Szalay estimator \citep{landyszalay} to first estimate $\xi_\mr{cc}$ as
\begin{equation}
\widehat{\xi}_{\rm cc}=\frac{\left <D D  \right >-2 \left <D R \right >+\left <R R \right >}{\left <R R \right >},\label{eq:landy-szalay}
\end{equation}
Where $DD$, $DR$, and $RR$ are the normalized number of cluster-cluster, cluster-random, and random-random pairs, respectively. For the randoms, we use the publicly available RM random catalog. The measured $\xi_\mr{cc}$ is then converted to $w_\mr{p}(R)$ using \Cref{eq:xicctowp}, with $\pi_\mr{max} = 100~ \mr{Mpc}/h$.

For our likelihood analyses, we use a jackknife resampling of our data vector with 83 independent regions to estimate our covariances. As in \Cref{sec:sim}, we assume a block-diagonal covariance matrix by treating the 9 sections of our data vector (abundances, 4 lensing profiles, and 4 autocorrelation functions) as independent from one another. We also apply the Hartlap correction \citep{hartlap} to perform a basic debiasing of our inverse covariance matrix used in the likelihood calculations. We show our measurements and their 1-$\sigma$ uncertainties in \Cref{fig:meas}. As the SDSS RM catalog and the associated observables we use have already been thoroughly studied and published, we do not consider a measurement level blinding for our analysis. Our blinding thus is limited to the analysis choice side, where we completely freeze the exact specifications of our fiducial analysis in \Cref{sec:sim:test} and apply it to both the validation in \Cref{sec:sim:val} and the real data analysis below.

\subsection{Results}

We show the results from our fiducial analysis on SDSS RM clusters in \Cref{fig:results_fid}. We consider this analysis blinded in the same way our validation was blinded, as we use the same frozen model from \Cref{sec:sim:test}, which was developed without any use of the SDSS data, for our results. Given the recent issues pointed out in \cite{dessamplers} that the \texttt{multinest} sampler may have trouble in obtaining the correct credible intervals for the sampled parameters, we tested both \texttt{multinest} and \texttt{polychord} samplers in sampling our posteriors. We find that while the two samplers yield very similar posterior centers, \texttt{polychord} yields wider credible intervals than \texttt{multinest} at the level of tens of per cents. We thus choose to use \texttt{polychord} for all of our real data results, which we consider as a conservative choice in line with the recommendations of \cite{dessamplers}.

\begin{figure}
    \includegraphics[width=\columnwidth]{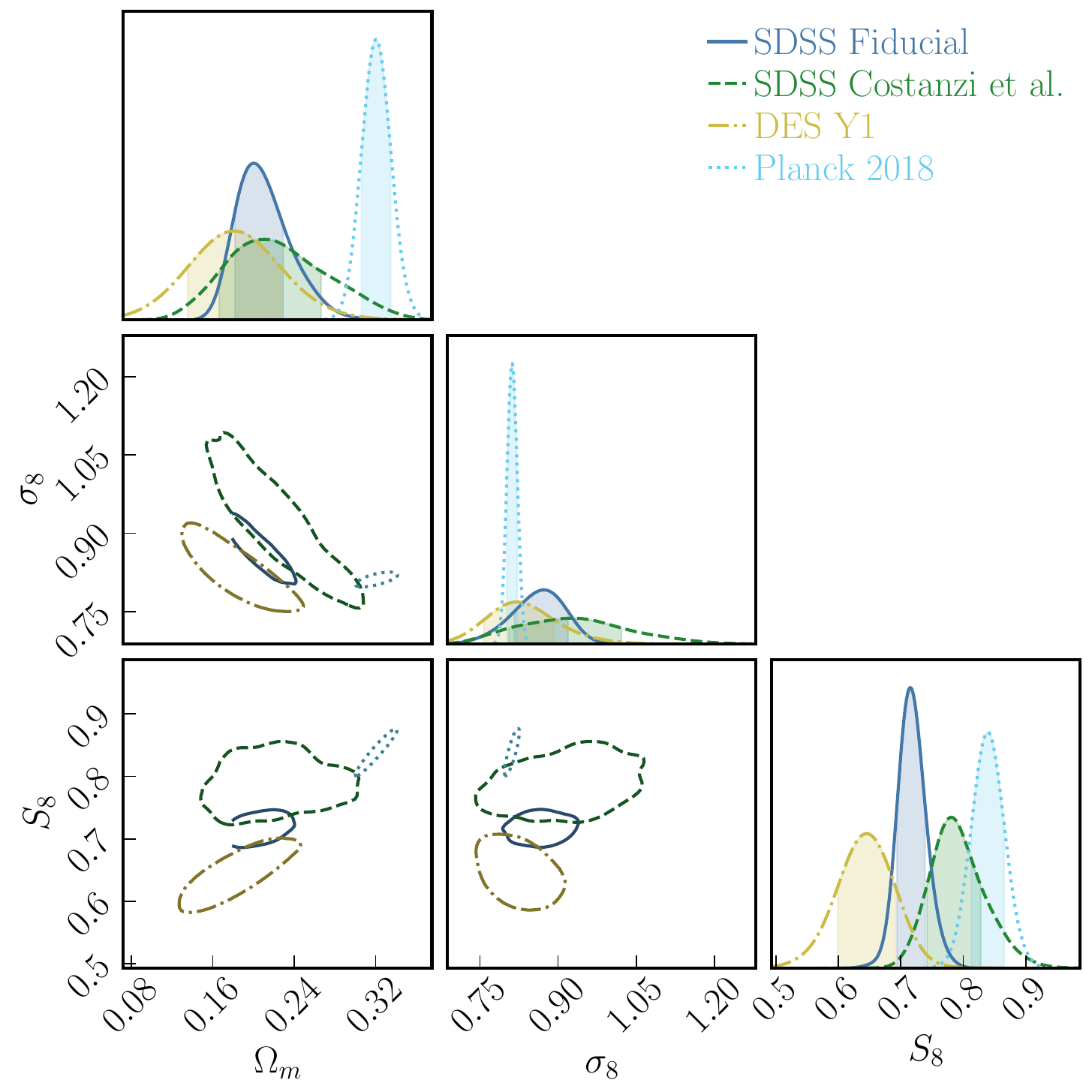}
    \caption{
    Comparison of 68\% confidence regions for cosmological parameters of interest from our fiducial 
    results (blue) with relevant previous results: \citet{costanzisdss} in green, \citet{desy1cl} in yellow, and 
    \citet{Planck18} in light blue.
    }
    \label{fig:results_fid_ext}
\end{figure}

Due to limits in the parameter space coverage of \texttt{darkemu}, in particular the lower limit of $\Omega_\mr{m} \gtrsim 0.18$, our posterior is truncated at around the 68\% contour on the low-$\Omega_\mr{m}$ end. This prevents us from robustly quantifying our cosmological posteriors, but the $S_8$ constraints are the least affected by the truncation and we report $S_8 = 0.715_{-0.021}^{+0.024}$. We also find that our posterior favors significantly lower $\Omega_\mr{m}$ and consequently $S_8$ values than other CMB/LSS results. \Cref{fig:results_fid_ext} further clarifies this tendency; compared against \textit{Planck} 2018 \citep{Planck18} results, we find significantly lower constraints for both $\Omega_\mr{m}$ and $S_8$. The rest of the parameter space shown consists of 4 MOR parameters ($A$, $B$, $\sigma_0$, and $q$), 2 miscentering parameters ($f_\mr{mis,1}$ and $R_\mr{mis,1}$), the photo-$z$ scatter ($\sigma_\mr{ph}$), and 3 anisotropic boost parameters ($c$, $R_{0,1}$, $\Pi_{0,1}$). For richness bin specific parameters (miscentering and anisotropic boost) we only show results for the first bin, but note that the results are consistent across all richness bins. The full constraints for these bin-specific parameters are shown in \Cref{fig:a1,fig:a2}. We find a best fit $\chi^2/\mr{dof}$ of $42.3/49$.

Let us take a closer look at the non-cosmology posteriors. First, we find that our systematics parameters are successfully self-calibrated. We find anisotropic boost amplitudes ranging between 15--20\%, which matches our findings from simulations as well as the observational constraints from \cite{tokrause2}. We also find miscentering fractions $\sim$ 15\% and miscentering scales around 0.2--0.3~Mpc$/h$, consistent with X-ray studies on the same cluster sample \citep[e.g.][]{miyatakecl,zhang18}. The degree of cluster photo-$z$ scatter is determined to be $\sim$ 20~Mpc$/h$, which again is consistent with the performance of the RM photo-$z$ estimation. Turning our attention to the degeneracies between the cosmological parameters and the rest of the parameter space, we find that beyond the MOR parameters, which are inherently correlated with cosmology, the only model parameter that shows a meaningful degeneracy with $\Omega_\mr{m}$ and $S_8$ is $c$. If we imagine extrapolating the degeneracy between $c$ and cosmology, $c\sim 0$ would result in cosmological constraints much closer to existing results. As presented in \Cref{eq:anisoboost}, $c$ represents the (generally negative) slope of the logarithmic decay in the effective bias $\Pi(R)$. A stronger (more negative) $c$ has the effect of suppressing the cluster clustering signal, while slightly enhancing the large scale regime of the cluster lensing signal. We investigate this degeneracy further in \Cref{sec:post}.

In addition, \Cref{fig:results_fid_ext} shows an interesting consistency between previous cluster cosmology results and ours. Both the \cite{costanzisdss} and the post-unblinding \cite{desy1cl} results center around lower $\Omega_\mr{m}$, and to some degree $S_8$, values, that are consistent with our findings. The differing approaches taken by these three cluster cosmology results make this consistency particularly intriguing. The analysis of \cite{costanzisdss} uses virtually the same abundance and lensing data as ours, coupled with a more complex mass-observable relation and systematics models calibrated from simulations. They do not, however, explicitly model the impact of anisotropic boosts on the lensing signal. \cite{desy1cl}, on the other hand, uses a completely independent data set, inherits the modeling of \cite{costanzisdss}, and calibrates the impact of anisotropic boosts on the weak lensing mass calibrations in their post-unblinding results using simulations. The two existing cosmology studies also vary the sum of neutrino masses in their analyses, while we fix it in our analysis. Considering these different combinations of models and data sets used, the consistently low $\Omega_\mr{m}$ found by all three results could suggest an unknown systematic effect or physics in clusters that cannot be localized to a particular data set or analysis method.

\begin{figure*}
	\includegraphics[width=2\columnwidth]{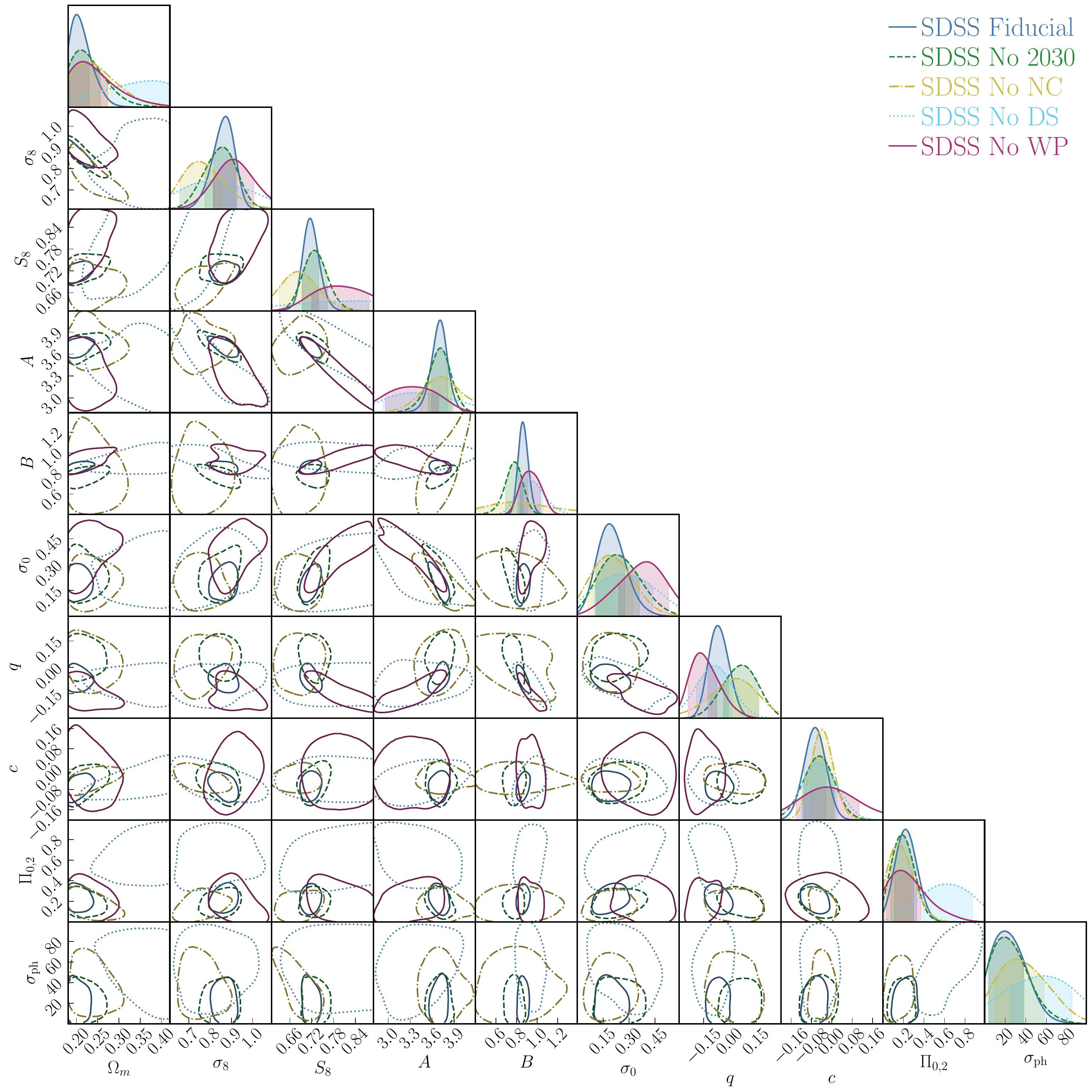}
    \caption{Comparison of 68\% confidence regions for model parameters when varying our analysis choices. Here we show the results from excluding the lowest richness bin 
    (``No 2030''), the abundance information (``No NC''), the lensing information (``No DS''), and the clustering information (``No WP''). We omit several parameters we show in \Cref{fig:results_fid}, as they do not show meaningful impacts on the cosmology results. Note that we have ``zoomed in'' toward the fiducial posterior center for readability.}
    \label{fig:results_fid_drop}
\end{figure*}

\section{Post-unblinding analyses}
\label{sec:post}

\subsection{Varying analysis choices}
\label{sec:post:drop}

\begin{figure*}
	\includegraphics[width=2\columnwidth]{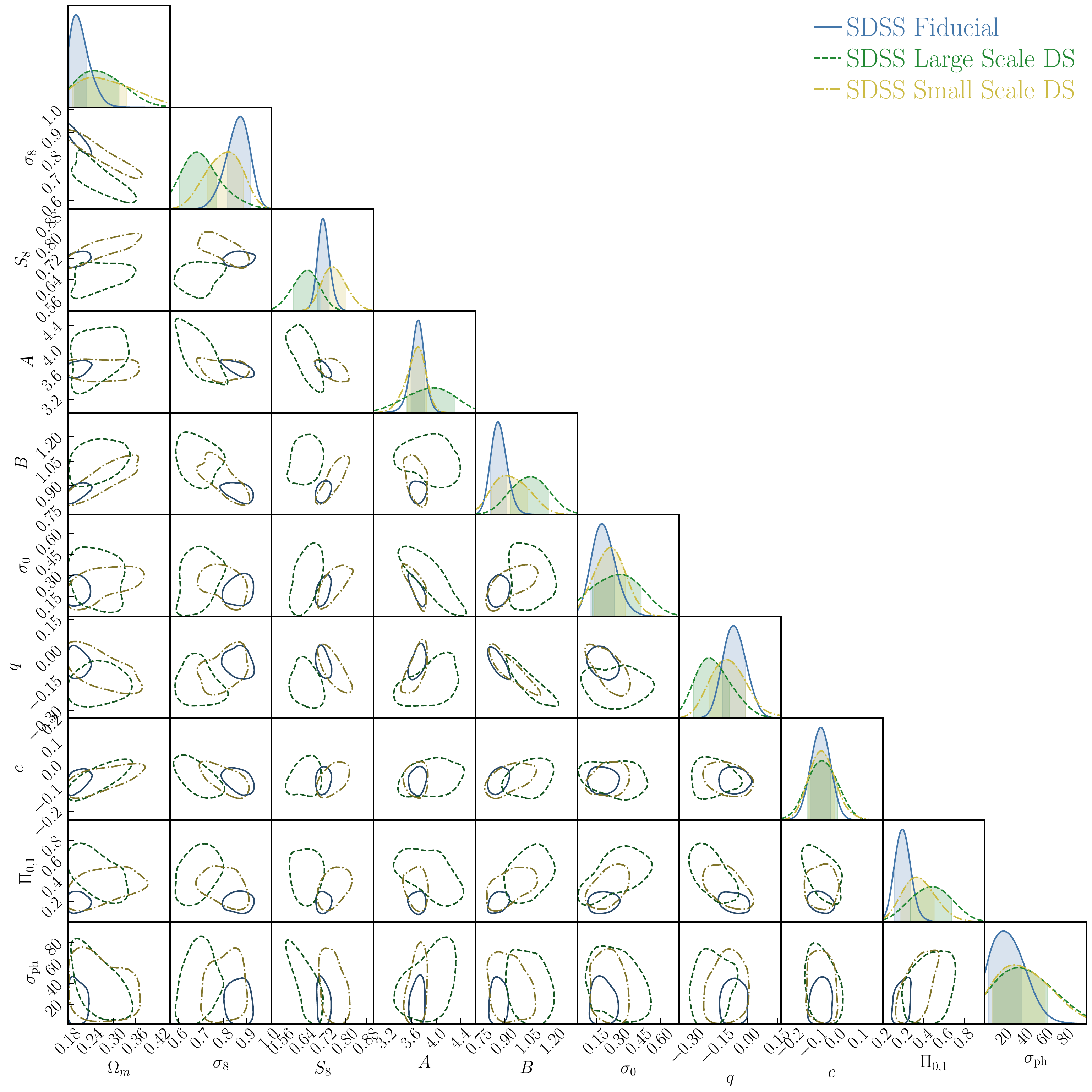}
    \caption{Similar to \Cref{fig:results_fid_drop}, but now varying the range of scales used for the cluster lensing signal. Here we show the 68\% confidence regions from our fiducial results (blue), as well as from restricting the range of scales for $\Delta\Sigma(R)$ to $R \geq 8~\mr{Mpc}/h$ (green) and  $R < 8~\mr{Mpc}/h$ (yellow). }
    \label{fig:results_fid_dsrange}
\end{figure*}

\begin{figure*}
	\includegraphics[width=2\columnwidth]{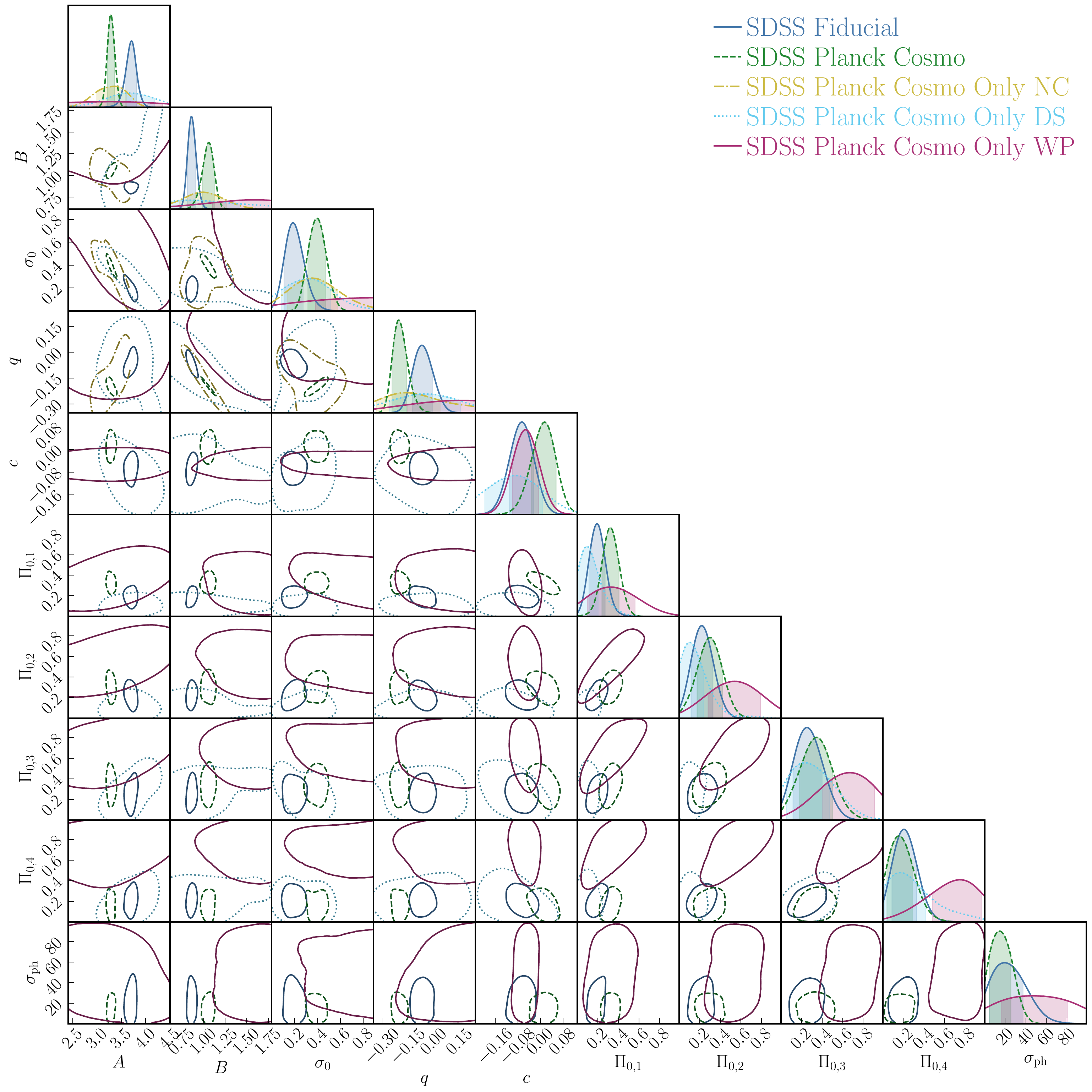}
    \caption{Comparison of 68\% confidence regions for MOR and anisotropic boost parameters when fixing the cosmology to the \textit{Planck} 2018 best fit and using individual sections of the data vector. We show fixed cosmology results using abundance information only (``Only NC''), lensing information only (``Only DS''), and clustering information only (``Only WP''). We also show the results from our fiducial analysis (blue; note that this does not assume a \textit{Planck} cosmology), as well as fixed cosmology results using the full data vector (green), for reference. Note that we have ``zoomed in'' toward the fiducial posterior center for readability, as we did in \Cref{fig:results_fid_drop}.}
    \label{fig:results_planck_onlys}
\end{figure*}

To better understand the driver behind our low-$\Omega_\mr{m}$ results, we perform a number of post-unblinding analyses. We begin with the argument in \cite{desy1cl} that their measured lensing signal, in particular for the lowest richness bin, is lower than expected. Following this insight, we test in \Cref{fig:results_fid_drop} whether dropping the lowest richness bin ($20\leq \lambda < 30$), or any one sector of our data vector (abundances, lensing, and clustering), impacts our conclusions. The results we find, however, are inconclusive with respect to the DES Y1 argument. While dropping the lowest richness bin leads to wider confidence regions, the posterior peaks for the cosmological parameters stay the same. Note that any potential broadening of uncertainties toward even lower $\Omega_\mr{m}$ would be truncated by our lower limit of $\Omega_\mr{m} \geq 0.18$. Dropping the entire lensing data does pull our results to higher $\Omega_\mr{m}$, but we also suffer a heavy loss in constraining power as the lensing information is our main cluster mass calibrator. Another notable aspect is that when we exclude WP, we find that our $S_8$ constraints shift higher. As a result, our ``No WP'' results become consistent with \cite{costanzisdss} in the $\Omega_\mr{m} - \sigma_8$ plane, and our marginalized $S_8$ constraints become consistent with other CMB/LSS based results. However, the low-$\Omega_\mr{m}$ issue is still persistent, and the posteriors are truncated, so we remain cautious in further interpreting this shift. We finally note that in plotting the posteriors we have ``zoomed in'' toward the posterior center of our fiducial result at the cost of cutting off some of the wider contours, in order to make the tighter confidence regions readable.

This exercise offers a number of insights on our fiducial results as well. First, we find that excluding any sector of the data vector leads to a significant loss of constraining power. While this is somewhat expected for NC or DS, it is interesting to see that excluding WP brings noticeable changes. This implies that cluster autocorrelation functions may be playing an important role in breaking the degeneracy among $\sigma_8$, cluster bias, and anisotropic boosts, all of which similarly affect the large-scale lensing and clustering amplitudes. Also, considering that results from simulations and observations, as well as results comparing galaxy autocorrelations against cluster-galaxy cross-correlations \citep{sunayama22} suggesting anisotropic boost strengths around 20\%, the ``No DS'' amplitudes of 60--70\% are not likely to be genuine. In addition, examining the $\Omega_\mr{m}-c$ and $\sigma_8-c$ planes clarify the driver behind the degeneracies among these parameters we found in our fiducial result. While every other result in \Cref{fig:results_fid_drop} shows similar degeneracies between $\Omega_\mr{m}/\sigma_8$ and $c$, the ``No WP'' result clearly departs from this tendency. This suggests that the clustering sector of our data, and more specifically the significant impact of $c$ on the predicted clustering amplitude, is correlating $c$ with the cosmological parameters of interest. As the ``No WP'' result also favors low values of $\Omega_\mathrm{m}$, it is difficult to argue that $c$, or its degeneracy with cosmological parameters, is responsible for our low $\Omega_\mathrm{m}$ results.

Noting that excluding lensing was the only variation we found to result in shifted cosmological constraints, we further test the lensing sector of our data vector in \Cref{fig:results_fid_dsrange}, where we limit the range of scales used in our cluster lensing signal to small scales only ($R < 8~\mathrm{Mpc}/h$) and large scales only ($R \geq 8~\mathrm{Mpc}/h$). Both variations result in posteriors favoring higher $\Omega_\mr{m}$, but similar to the ``No DS'' case these results accompany changes in the MOR and anisotropic boost parameter constraints. When only the small scale lensing data is used, parameter constraints show a broadening of the constraints along the degeneracy directions, and the $S_8$ constraints shift higher. As for the large scale lensing only case, the changes have different characteristics, and involve larger departures from the fiducial result in MOR and anisotropic boost parameters as well as even lower values of $S_8$. Limiting ourselves to the cosmological parameters, we find that both the ``Small Scale DS'' and ``Large Scale DS'' results are consistent with the ``No DS'' results, with the former slightly more so than the latter. Note that the small scale lensing signal probes cluster masses with a higher signal-to-noise, as well as a weaker contribution from anisotropic boosts, compared to its large scale counterpart. This explains the wider confidence regions seen from the ``Large Scale DS'' results: without the small scale lensing information our model finds it more difficult to constrain parameters related to mass calibration.

We end this section by cautioning the reader that all of the results presented in \Cref{fig:results_fid_drop,fig:results_fid_dsrange} are correlated with one another, as any given pair of the results shown shares in common at least one sector of the data vector. This, in addition to the fact that our cosmology posteriors are necessarily truncated at the parameter limits of the Dark Emulator, prohibits straightforward quantifications of differences, or tensions, between these results. We thus leave the task of formally estimating the statistical significance of these tensions as future work.

\begin{figure*}
	\includegraphics[width=1.9\columnwidth]{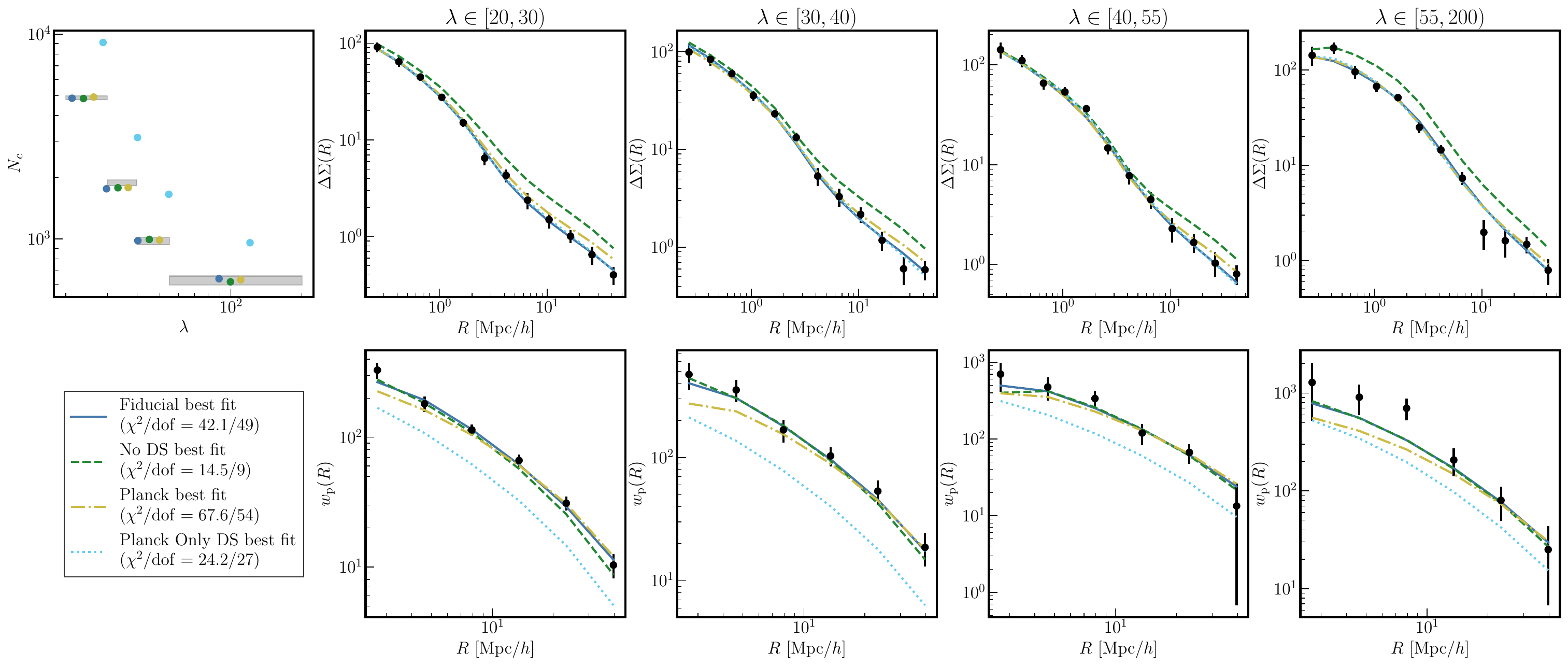}
    \caption{Comparison of our measurements with the best-fit predictions from the fiducial analysis and selected post-unblinding tests. We show cluster abundance comparisons in the top left panel, and the cluster lensing (top row) and cluster clustering (bottom row) comparisons in the next four columns, with each column corresponding to a richness bin we analyze. The bests fits shown are from the fiducial analysis (blue), the ``No DS'' analysis (green), full data analysis under \textit{Planck} cosmology (yellow), and the ``Only DS'' analysis under \textit{Planck} cosmology (light blue).}
    \label{fig:measpred}
\end{figure*}

\subsection{Fixing cosmological parameters}
The results in \Cref{sec:post:drop} do not offer a straightforward interpretation. Removing lensing from the analyzed data shifts cosmological constraints closer to existing results, but the surprisingly high strength of anisotropic boosts predicted makes it difficult to take at face value. Using subsets of the lensing data does not produce a clear picture either; using only large scale lensing yields even lower values of $S_8$, and it is difficult to tell whether the small scale lensing only results are exhibiting genuine shifts in cosmological constraints or a mere broadening of the contours. At the same time, no other exclusion of data vector sectors or richness bins seem to have a significant effect on the low values of our marginalized $\Omega_\mr{m}$ constraints. To decouple the different elements at play, we now fix our cosmology to the \textit{Planck} 2018 best fit and consider only the non-cosmological parameter space.

As fixing the cosmology significantly shrinks the parameter space and the degeneracies therein, we can take a step further from \Cref{sec:post:drop} and derive constraints using individual sectors of our data vector, in addition to using the full data set combination. \Cref{fig:results_planck_onlys} shows the results from this exercise, and this further illuminates how each sector drives the MOR and anisotropic boost constraints. Note that we are again ``zooming in'' for readability, as we did above in \Cref{fig:results_fid_drop}. Starting with the full data analysis under fixed cosmology, we first note that the results favor an MOR that is clearly different from the fiducial results. More specifically, we see the new preferred MOR has a lower normalization ($A$), a steeper slope ($B$), a larger intrinsic scatter ($\sigma_0$), and a stronger mass-dependent scatter ($q$). We also find shifts in the anisotropic boost constraints, with the fixed cosmology results preferring (i) $c \sim 0$ and (ii) slightly larger, and richness dependent, boost amplitudes. The richness dependence of the $\Pi_{0,i}$ parameters was not apparent in any of our earlier results.

Turning our attention now to single sector constraints, we immediate notice that the ``Only NC'' result shows a clear bimodality in its MOR constraints. Interestingly, the two modes of this posterior closely follow the respective preferred MORs of the fiducial and the fixed cosmology full data analyses. It is surprising to see that the fiducial MOR is still one of two preferred modes, as this was derived under a very different cosmology from the one assume here. The ``Only DS'' results also show MOR constraints compatible with both of these modes, and the ``Only WP'' MOR constraints are too wide to interpret with significance. As for the anisotropic boost parameters, we note that the ``Only WP'' results favor boost strengths that are significantly higher and even more richness dependent than the full analysis. The higher strengths are in line with the ``No DS'' results from \Cref{fig:results_fid_drop}, and may be related to fact that the photo-$z$ scatter, largely unconstrained and thus generally higher than the full analysis best fit, ends up suppressing the predicted WP signals more. The richness dependence, on the other hand, may suggest that the bias-mass relation dictated by the Planck cosmology, combined with our model, is not compatible with the measured amplitudes of the cluster autocorrelation functions. The ``Only DS'' results, on the other hand, prefer weaker boost strengths, also weakly richness dependent, but are generally consistent with the full analysis results.

Let us consolidate our findings in the data space with \Cref{fig:measpred}. Here we compare our measurements with the best fit predictions from our fiducial analysis and a number of relevant post-unblinding tests. We focus in particular on the DS sector, as the ``No DS'' result was the only result we saw with any shift in the $\Omega_\mathrm{m}$ posterior center. Beginning with the fiducial best fit, we find good agreement across all sectors between measurements and predictions. When lensing is removed from this analysis, the preferred cosmology becomes consistent with \textit{Planck} and the MOR also shifts. The resulting predictions, shown in green, fit the measurements included in the analysis (NC and WP) well but overpredicts lensing, in particular on large scales. Note that since the ``No DS'' fits do not include miscentering parameters, we use their best fit values from the fiducial analysis to generate lensing predictions. Next, with the cosmology fixed to the \textit{Planck} 2018 values, the resulting best fit again slightly overpredicts lensing on large scales, and this time the clustering signal is underpredicted. This ``compromise'' between the lensing and clustering sectors, compared to the  fiducial ``No DS'' prediction, is likely driven by the fact that lensing dominates in the signal-to-noise over clustering. In addition, the degradation in the goodness-of-fit, from $\chi^2/\mr{dof}$ of $42.1/49$ to $67.6/54$, illustrates potential internal tensions between the different data sectors when assuming a \textit{Planck} cosmology.

Finally, when we assume the \textit{Planck} cosmology and fit only the lensing data, we note significant overpredictions in the cluster abundances and underpredictions in the clustering.  This presents the clearest message among all of the post-unblinding tests that we have performed: under a \textit{Planck} cosmology, our lensing data vector underpredicts cluster masses, leading to overpredicted abundances and underpredicted clustering amplitudes. Given the demonstrated completeness and purity of SDSS RM clusters, as well as the consistency of the mispredictions between the abundance and clustering sectors, the alternative interpretation --- abundance measurements suppressed by a factor of two and clustering measurements inflated by a factor of two --- is unlikely. The possibility of suppressed cluster lensing signals in our data is particular notable, as this is in line with findings from the DES Y1 cluster cosmology analysis. Considered together, we arrive at a stronger case for a significant missing element in our current understanding of cluster lensing.

\section{Summary and Discussion}
\label{sec:conc}
In this paper, we discussed the development and validation of a novel cluster cosmology analysis designed to be robust against the anisotropic boosts impacting the cluster lensing and cluster clustering observables, and the application of this analysis  on the SDSS RM catalog to obtain cosmological constraints. Our work and findings can be summarized as follows.
\begin{itemize}
    \item We accounted for the anisotropic boosts in the lensing and clustering signals around clusters with an empirical model taking the form of a scale-dependent effective bias. Upon validation with a blind cosmology challenge on mocks, this model was able to recover the expected degrees of anisotropic boosts on the lensing and clustering signals.
    \item We in addition find from our validation that by jointly analyzing cluster abundances, cluster lensing, and cluster clustering, especially across a wide range of scales using fully non-linear predictions from \texttt{darkemu}, we are able to obtain tight and unbiased constraints on the cosmological parameters of interest --- $\Omega_\mr{m}$ and $\sigma_8$ --- even with flat uninformative priors on all model parameters.
    \item After applying the pipeline to the SDSS RM catalog, we find cosmological constraints that strongly favor low $\Omega_\mr{m}$ and to a lesser degree $S_8$ values. We note that these results are interestingly consistent with recent attempts at cosmological constraints using clusters, but in strong tension with other CMB/LSS results.
    \item From a series of post-unblinding analyses, we find hints for internal tensions among the different sectors of our data set. Our most notable finding from these tests is that the cluster lensing signal we measure significantly underpredicts cluster masses when assuming a \textit{Planck} cosmology. We are actively investigating the nature and origin of these findings.
\end{itemize}

The fact that our fiducial pipeline performed well on a blind challenge on mocks, but obtained somewhat suspicious (albeit with confirmation bias) cosmology results on real data, allows for informed guesses on the driver behind our cosmological constraints. Specifically, we may postulate that this driver is a subset of the differences between the mocks and the real data. Regarding the cluster lensing signal, the most likely driver behind our cosmology results, we note the difference in the way lensing is measured. In the mocks, we have the true underlying matter distribution, which we can readily project along the line-of-sight to calculate the lensing observable $\Delta\Sigma(R)$. In real data, however, we infer this quantity through the ellipticities of background galaxies, and effects that perturb the connection between ellipticities and matter densities, e.g., unmodeled intrinsic alignments, can complicate our findings. Such effects would cross boundaries of data sets and modeling approaches, and it would be very significant for cluster cosmology if we could point to a single cause that can explain the consistent disagreement between multiple cluster cosmology results and existing CMB/LSS cosmology constraints. The use of shape measurements from the Hyper Suprime-Cam SSP (HSC) survey \citep{hscy3} may become particularly useful in investigating these effects. The HSC data overlaps with the SDSS RM footprint, but has significantly deeper photometry and better image quality than the SDSS source catalog, enabling source galaxy selections more robust to effects such as intrinsic alignments or source-cluster member confusion.

A deeper look into the characteristics of our model and data is also warranted. For instance, in modeling the impact of the anisotropic boosts we assume that the boost in clustering is the square of the boost in lensing, and carry it on all scales down to 2~Mpc$/h$. While this assumption has been successfully validated on mocks, we also know that tracer biases can deviate from this linear ansatz on small scales. It may be interesting to consider more flexibility on small scales for the anisotropic boost models, or to craft an entirely new approach based on an explicit model for the anisotropic LSS surrounding optical clusters. In addition, the flat uninformative priors we employ throughout our analysis may give rise to undetected parameter volume effects. Again, while we have not seen this in our validation, further checks on the real data results may bring surprises. We plan to explore all of the above possibilities in greater depth with follow-up studies in the near future, and we remain optimistic that these efforts will culminate in a robust analysis of clusters for precision cosmology.

\section*{Acknowledgements}
We thank Rachel Mandelbaum for providing us the SDSS shape catalogs critical for our cluster lensing measurements, and Matteo Costanzi for providing us the posterior samples for the \cite{costanzisdss} and \cite{desy1cl} results. They have also provided useful comments on earlier drafts of this work. We also thank Elisabeth Krause and Eduardo Rozo for helpful discussions on our post-unblinding tests.

This work was supported in part by the World Premier International Research Center Initiative, MEXT, Japan; JSPS KAKENHI Grants No.~JP17K14273, JP19H00677, JP20H05850, JP20H05855,  and JP21H01081; the Japan Science and Technology Agency (JST) CREST JPMHCR1414; the JST AIP Acceleration Research Grant Number JP20317829; and the Basic Research Grant (Super AI) of Institute for AI and Beyond of the University of Tokyo. The numerical simulations were performed on Cray XC50 at Center for Computational Astrophysics, National Astronomical Observatory of Japan.
Additionally, 
YP was supported by JSPS KAKENHI Grant Number JP21K13917;
TS was supported by the Grant-in-Aid for JSPS
Fellows 20J01600;
YK was supported by the Advanced Leading Graduate Course for Photon Science at the University of Tokyo;
HM was supported by JSPS KAKENHI Grant Number JP20H01932;
SS was supported by JSPS KAKENHI Grant Number 21J10314 and by the International Graduate Program for Excellence in Earth-Space Science (IGPEES), World-leading Innovative Graduate Study (WINGS) Program, the University of Tokyo.

Funding for SDSS-III has been provided by the Alfred P. Sloan Foundation, the Participating Institutions, the National Science Foundation, and the U.S. Department of Energy Office of Science. The SDSS-III web site is \url{http://www.sdss3.org/.}

SDSS-III is managed by the Astrophysical Research Consortium for the Participating Institutions of the SDSS-III Collaboration including the University of Arizona, the Brazilian Participation Group, Brookhaven National Laboratory, Carnegie Mellon University, University of Florida, the French Participation Group, the German Participation Group, Harvard University, the Instituto de Astrofisica de Canarias, the Michigan State/Notre Dame/JINA Participation Group, Johns Hopkins University, Lawrence Berkeley National Laboratory, Max Planck Institute for Astrophysics, Max Planck Institute for Extraterrestrial Physics, New Mexico State University, New York University, Ohio State University, Pennsylvania State University, University of Portsmouth, Princeton University, the Spanish Participation Group, University of Tokyo, University of Utah, Vanderbilt University, University of Virginia, University of Washington, and Yale University.

\section*{Data Availability}

The SDSS DR8 redMaPPer cluster and cluster member catalogs are publicly available at \url{http://risa.stanford.edu/redmapper}. The SDSS shape catalog used in this work is not publicly available; please contact Rachel Mandelbaum for inquiries on obtaining access.



\bibliographystyle{mnras}
\bibliography{refs} 




\appendix

\section{Full constraints from the Blind Validation}
\label{app:a}
\begin{figure*}
	\includegraphics[width=2\columnwidth]{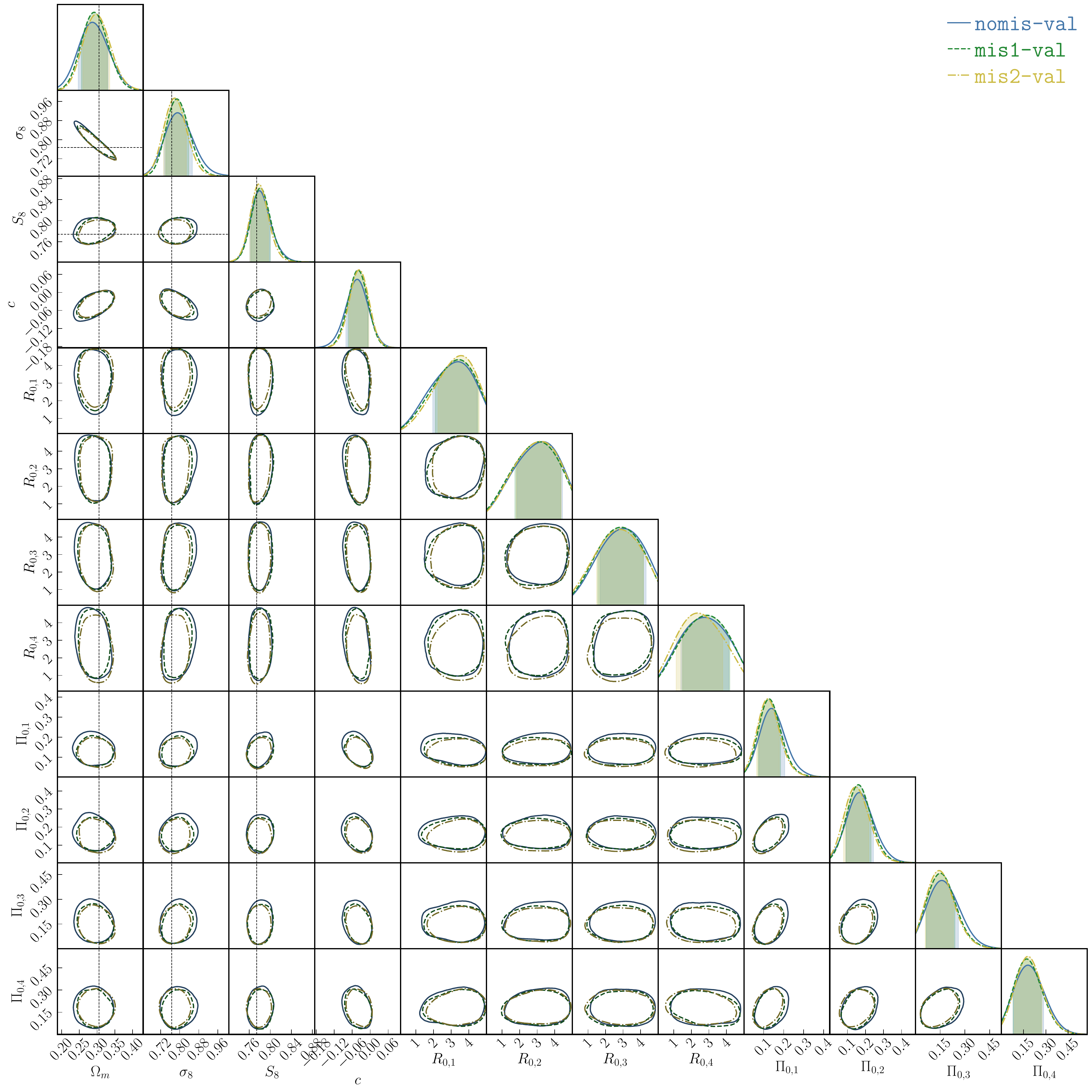}
    \caption{The 68\% confidence regions for the full set of anisotropic boost parameters, along with the cosmological parameters of interest, from our validation analyses. We show results from all three variants of our validation test: \texttt{nomis-val}, \texttt{mis1-val}, and \texttt{mis2-val}.}
    \label{fig:val-boost}
\end{figure*}

\begin{figure*}
	\includegraphics[width=2\columnwidth]{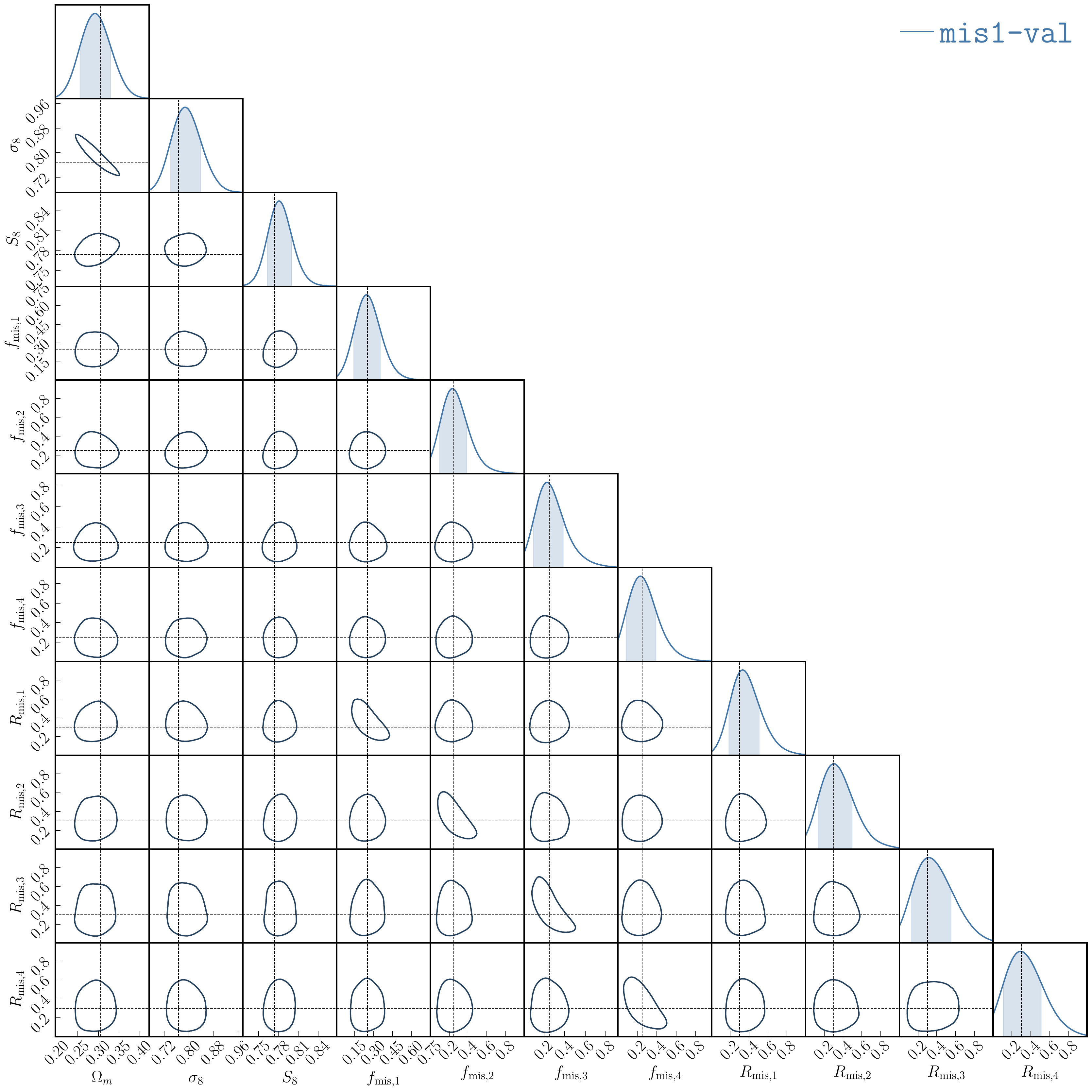}
    \caption{The 68\% confidence regions for the full set of miscentering parameters, along with the cosmological parameters of interest, from our validation analysis. We only show the \texttt{mis1-val} results here, where the input and analysis miscentering models match.}
    \label{fig:val-mis}
\end{figure*}

We show here the full constraints from our validation analysis. \Cref{fig:val-boost,fig:val-mis} respectively show the anisotropic boost and miscentering constraints we obtain, along with the cosmological parameters of interest. Note that for the latter we have the true input parameter values available, and we limit the constraints shown to the \texttt{mis1-val} case, where the input and analysis miscentering models match.

\section{Full constraints from the Fiducial SDSS RM Analysis}
\label{app:b}
\begin{figure*}
	\includegraphics[width=2\columnwidth]{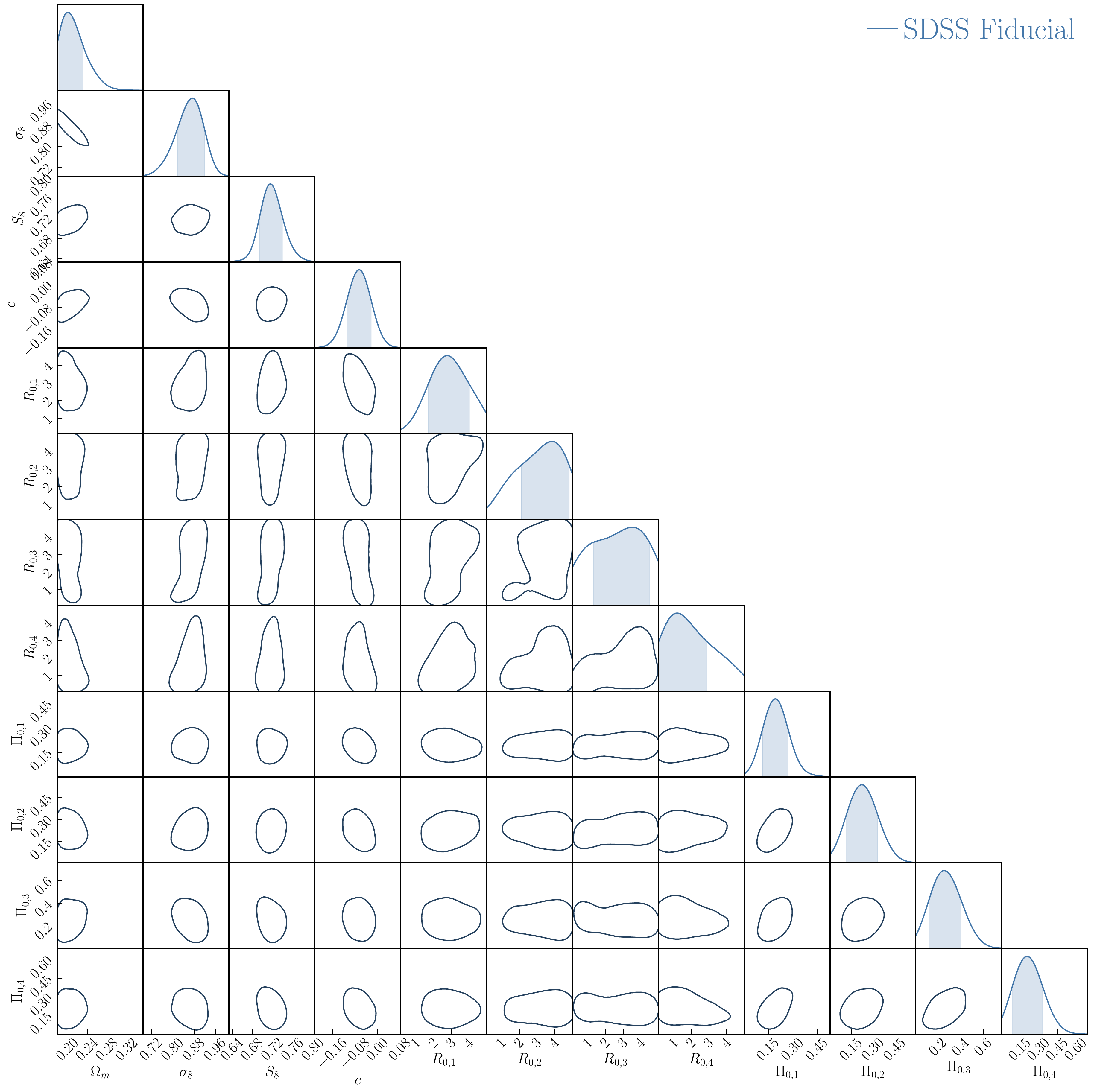}
    \caption{The 68\% confidence regions for the full set of anisotropic boost parameters, along with the cosmological parameters of interest, from our fiducial analysis.}
    \label{fig:a1}
\end{figure*}

\begin{figure*}
	\includegraphics[width=2\columnwidth]{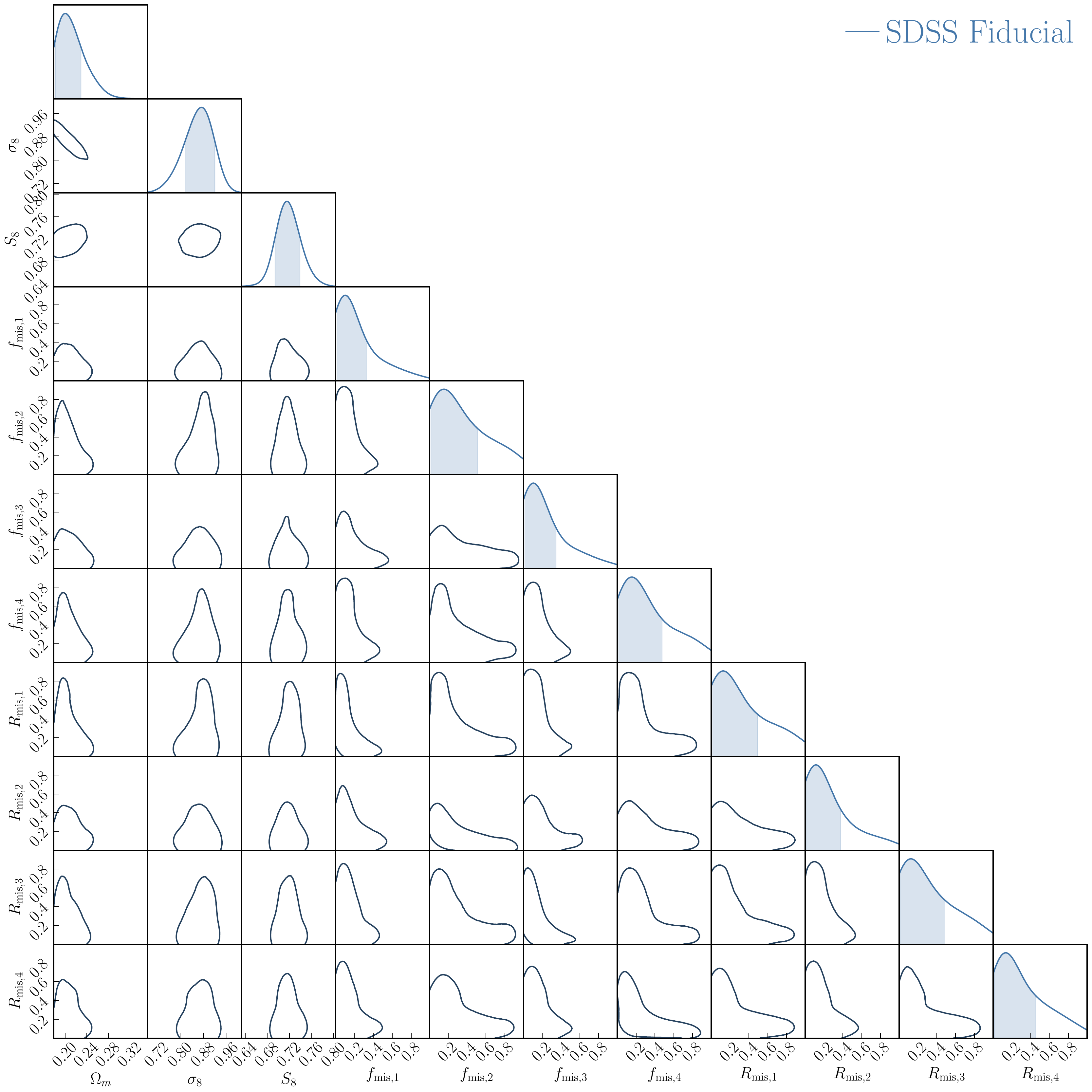}
    \caption{The 68\% confidence regions for the full set of miscentering parameters, along with the cosmological parameters of interest, from our fiducial analysis.}
    \label{fig:a2}
\end{figure*}

We show here the full constraints from our fiducial analysis for the bin-specific parameters. \Cref{fig:a1,fig:a2} respectively show these results for the anisotropic boost and miscentering parameters, along with the cosmological parameters of interest. We note that the anisotropic boost and miscentering constraints show consistent behavior across all richness bins, despite being freely varied for each of these bins.

\bsp	
\label{lastpage}
\end{document}